\documentclass[aps,prl,preprintnumbers,amsmath,amssymb,latexsym,nofootinbib,array,enumerate,letter,twocolumn,superscriptaddress]{revtex4-1}

\usepackage{here}
\usepackage{comment,braket}
\usepackage{epsf}
\usepackage{amsmath}
\usepackage{graphics}
\usepackage{amsfonts}
\usepackage{amssymb}
\usepackage{latexsym}
\usepackage{color}
\usepackage{ulem}
\usepackage{feynmf}
\usepackage{enumerate}
\usepackage{lipsum}
\usepackage[pdftex]{graphicx}
\usepackage{bm}
\usepackage[pdftex]{hyperref}
\usepackage[svgnames]{xcolor}
\definecolor{phthaloblue}{rgb}{0.0, 0.06, 0.54}
\hypersetup{
    colorlinks=true,
    linkcolor=phthaloblue,
    citecolor=blue,
    filecolor=blue,
    urlcolor=phthaloblue,
    }

\newcommand{\GeV}{\  {\rm GeV} }
\newcommand{\TeV}{\  {\rm TeV} }
\newcommand{\pr}[1]{\left(#1\right)}
\newcommand{\deriv}{{\rm d}}
\newcommand{\imwidth}{\columnwidth}
\newcommand{\micromegas}{\code{MicrOMEGAs5.0}~}
\newcommand{\feynrules}{\code{FeynRules}~}
\newcommand{\code}[1]{\texttt{#1}}
\newcommand{\eEDM}{e-EDM~}
\newcommand{\lagr}{\mathcal{L}}

\begin{document}
    
\title{The Race to Find Split Higgsino Dark Matter}

\author{Raymond T. Co}
\affiliation{Leinweber Center for Theoretical Physics, University of Michigan, Ann Arbor, MI 48109, USA}
\affiliation{William I. Fine Theoretical Physics Institute, School of Physics and Astronomy, University of Minnesota, Minneapolis, MN 55455, USA}
\author{Benjamin Sheff}
\affiliation{Leinweber Center for Theoretical Physics, University of Michigan, Ann Arbor, MI 48109, USA}
\author{James D. Wells}
\affiliation{Leinweber Center for Theoretical Physics, University of Michigan, Ann Arbor, MI 48109, USA}
\date{\today}

\begin{abstract}
Split higgsinos are a compelling class of models to explain dark matter and may be on the verge of detection by multiple current experimental avenues. The idea is based on a large split in scales between the electroweak scale and decoupled scalars, with relatively light higgsinos between the two. Such models enjoy the merit of depending on very few parameters while still explaining gauge coupling unification, dark matter, and most of the hierarchy between the Planck and electroweak scales, and they remain undetected by past experiments. We analyze split higgsinos in view of current and next generation experiments. We discuss the direct and indirect detection prospects and further demonstrate promising discovery potentials in the upcoming electron electric dipole moment experiments. The parameter space of this model is analyzed in terms of experiments expected to run in the coming years and where we should be looking for the next potential discoveries.
\end{abstract}

\preprint{LCTP-21-11, UMN-TH-4016/21, FTPI-MINN-21/09}

\maketitle

\section{Introduction}
The existence and prevalence of dark matter (DM) remains one of the most glaring gaps in the modern understanding of physics~\cite{Zwicky_DM_1933, Kolb_Turner}. One compelling class of models to explain this issue is a set of supersymmetric (SUSY) models in which the higgsino is the lightest superpartner (LSP), with high scale scalars and gaugino masses between the two. We will refer to the higgsino in this class of models as a split higgsino. Split higgsinos can come with many of the advantages of a SUSY model, such as gauge coupling unification and an explanation for the scale separation between the Planck and electroweak energies. The higgsino itself in these models is a form of weakly interacting massive particle (WIMP) that can be thermally produced in the early universe and is as yet inaccessible to experiment. Many SUSY models like the minimal supersymmetric Standard Model (MSSM) face significant constraints from the following:
\begin{itemize}
    \item Colliders place particularly strong limits on new physics with colored particles, and can provide generic limits on new physics up to a scale of order a few hundred GeV. A number of detailed reviews on the topic exist~\cite{Alves_Simplified_2012,Canepa_Searches_2019,Baer:2020kwz}.
    \item Flavor Changing Neutral Current (FCNC) experiments place precise limits on new flavor physics not protected by some version of the Glashow–Iliopoulos–Maiani mechanism~\cite{Gaillard_Rare_1974, Ellis_Flavor_1981}. Such constructions are not generically expected, and imply harsh constraints on SUSY models~\cite{Donoghue_Flavor_1983, Hall:1985dx, Baer_Landscape_2019}. For a detailed analysis of the topic, also see Ref.~\cite{Gabbiani:1996hi}.
    \item The stability of the proton places severe limits on dimension $> 4$ operators from new physics, which in turn place limits through effective theories from SUSY extensions~\cite{Weinberg_Supersymmetry_1982, Sakai_Proton_1981, Chamoun_Nucleon_2020}.
    \item Electric dipole moment (EDM) measurements limit the CP violation of new physics, which can be generically large in SUSY models, with large EDMs particularly arising from one-loop corrections involving scalar superpartners~\cite{delAguila:1983dfr, cesarotti_interpreting_2019}.
\end{itemize}
However, models with decoupled scalars remain open and largely untested, as the above constraints largely depend on scalar superpartners involved in the requisite interactions. Previous studies have analyzed similar models with various forms of neutralino DM~\cite{wells_implications_2003,arkani-hamed_well-tempered_2006, Baer:2016ucr, kowalska_discreet_2018}, but we will focus on the case of split higgsinos, a particularly simple model with the potential for near future discovery as we will detail in this paper.

Split higgsino models have a  nearly pure higgsino LSP, with gauginos being somewhat heavier and all the non Standard Model scalars having decoupled. If we assume standard cosmology and DM generation through thermal freeze-out, as per the usual story for WIMPs~\cite{Kolb_Turner}, higgsinos comprise all the DM if the mass is $m_{\rm DM}\sim 1.2$ TeV~\cite{Profumo_Statistical_2004,giudice_split_2005}. Other values are possible but require non-standard cosmologies. While many of the results discussed here will be generic, we will focus in particular on the compelling case of anomaly mediated SUSY breaking~\cite{randall_out_1999,Giudice:1998xp} generating the gaugino masses, reducing our model space further, as discussed in Sec.~\ref{sec:model}.

While split higgsinos are difficult to detect by more traditional approaches, such as colliders (see, e.g.\ \cite{Aaboud:2017leg}) and direct detection experiments, they can still be accessible to next generation indirect detection and electron electric dipole moment (e-EDM) experiments, as we will detail in this paper. In fact, we currently appear to be at the verge of reaching this theory space with real data.

\section{Model details}
\label{sec:model}

The main ingredients in a split higgsino model are a higgsino LSP and heavy scalars from SUSY breaking at a high scale~\cite{wells_implications_2003,giudice_split_2005,arkani-hamed_aspects_2005,arkani-hamed_well-tempered_2006}. While split higgsinos may require some degree of tuning in the Higgs potential (however, see e.g.~\cite{Baer:2012uy, Baer:2012up}), decoupling the scalars at a high, nearly degenerate mass wipes clean the bullet points discussed in the introduction, and simplifies the vast space of SUSY breaking parameters by decoupling many of them. The remaining masses, those of the gaugino and higgsino, lie between the electroweak scale and the scalar mass, forming the relevant new physics for phenomenological study.

As the scalars decouple at low scales, we get an effective Lagrangian
$\lagr = \lagr_{\rm SM} + \lagr_{\rm eff} $
for, up to kinetic terms,
\begin{align}
    \label{eq:model}
    -\lagr_{\rm eff} = & \ \frac{M_2}{2}\Tilde{W}^a\Tilde{W}^a+\frac{M_1}{2}\Tilde{B}\Tilde{B} + \mu \Tilde{H}_u\epsilon\Tilde{H}_d \\ \nonumber
    & + \frac{H^\dag}{\sqrt{2}}\pr{\Tilde{g}_u\sigma^a\Tilde{W}^a+\Tilde{g}'_u\Tilde{B}}\Tilde{H}_u \\
    & +  \frac{H^T\epsilon}{\sqrt{2}}\pr{\Tilde{g}_d\sigma^a\Tilde{W}^a+\Tilde{g}'_d\Tilde{B}}\Tilde{H}_d + h.c. . \nonumber
\end{align}
The couplings $\Tilde{g}_{u,d}$, when run up to the scalar mass scale, are $g\sin{\beta}$ and $g\cos{\beta}$ respectively,~\cite{giudice_split_2005} where $\tan\beta$ is the ratio of the vacuum expectation values of the up- and down-type Higgs, $H_u, H_d$, in MSSM. 
The couplings $\Tilde{g}_{u,d}'$ are similarly defined, but for hypercharge $g'$. The parameter $\mu$ comes from the superpotential, but here is effectively the higgsino mass. The fields $\Tilde{W}^a, \Tilde{B}, \Tilde{H}_u, \Tilde{H}_d$ are the wino, bino, and two higgsino gauge eigenstates respectively, while $H$ is the Higgs. Lastly, $\sigma^a$ are the Pauli matrix three vectors, and $\epsilon$ is the Levi-Civita tensor.

While there are a number of ways to construct split higgsino models, for illustration we consider the most natural approach, beginning with adding a charged, chiral supermultiplet $\mathcal{S} = S+\sqrt{2}\psi\theta + F_S\theta^2$ to the theory. The supermultiplet couples to the scalar superpartners, giving them a SUSY breaking mass of order $F_S/M_{Pl}$, which coincides with the gravitino mass $m_{3/2}$. A large value of $F_S$ gives large, nearly degenerate masses to the scalars. In particular, $m_{3/2}\sim \mathcal{O}$(PeV) is favored, as it generates the correct Higgs boson mass~\cite{ArkaniHame_supersymmetric_2005, Arbey_Implication_2012}, can allow full unification of gauge couplings at high energies, and can generate appropriate neutrino masses by coupling $\mathcal{S}$ to neutrinos and the Higgs boson~\cite{wells_pev-scale_2005}. Gauge invariance protects the gauginos and higgsinos, which gain mass by a conformal anomaly and a free parameter respectively. The SUSY breaking masses, $M_1, M_2, M_3$, for the bino, wino, and gluino respectively, are thereby suppressed from the scalar mass by a loop factor, arising from mediating a conformal anomaly. The resulting masses have a fairly strict relationship, $M_3 \simeq 10M_2 \simeq 3M_1$~\cite{randall_out_1999}, following the ratio of the respective gauge coupling beta functions.

The resultant model is compelling because it relies on very few fitted parameters, while providing a DM candidate and retaining most of the benefits of a supersymmetric model beyond admitting some limited tuning on the electroweak scale. Taking PeV scale scalars, we arrive at a $M_2 \sim$ 3-10 TeV, with the other gaugino masses set accordingly. The higgsino mass is a free parameter, but for the freeze-out WIMP scenario in a standard cosmology, it must have mass $\sim 1.2$ TeV to account for the observed dark matter abundance.

We expand our model space beyond this point to include values of $|\mu|$ from $10^2$ to $10^4$ GeV, and $M_2$ between $10^2$ and $10^5$ GeV, taking $|\mu| < M_2$, to form a broad analysis of split higgsino models. We will also largely ignore low values of $|\mu|$ due to pressures from experiments like colliders that place limits up to $|\mu| \sim \mathcal{O}(100)$ GeV.

We analytically calculate an electric dipole moment following the formalism in Ref.~\cite{giudice_electric_2006}. This property can arise due to complex phases in $M_1, M_2, \Tilde{g}_{u,d}, \Tilde{g}'_{u,d}$, or $\mu$, though through field redefinitions these are all equivalent to phases in $M_1$ and $M_2$. The \eEDM is compared against recent and projected experiments for limits and expected discovery reach.

The neutralino mass eigenstates differ from the gaugino and higgsino states shown in Eq.~(\ref{eq:model}) due to mixing whose size is based on the relative scales of $\mu$ and $M_i$. While we focus on the regime in which the lightest neutralino is a nearly pure higgsino, we will label mass eigenstates as $\Tilde{\chi}_i$ and $\Tilde{\chi}^\pm$ for the neutralinos and charginos respectively. While the mixing matrix for neutralinos could not be solved analytically, analytic approximations can be made for $M_i, \mu \gg m_Z$. This calculation is then entered into \feynrules\cite{feynrules}, which is used to generate model files for use in \micromegas\cite{micromegas}. \micromegas is then used to calculate annihilation and scattering cross sections for the DM higgsino in our model. These cross sections are compared to existing limits and projected discovery reaches for future experiments in the next two sections.

\section{Direct Detection}

Direct detection, detecting DM through scattering against a target material on Earth, is one of the most generic techniques in the search for WIMP DM. The leading limit comes from Xenon1T~\cite{aprile_dark_2018}, which uses 2 tons of liquid xenon as a target, detecting events based on scintillation and ionization from collisions between the DM and the xenon. By observing a total of 14 months, this experiment places significant limits on the product of the cross section and velocity for DM scattering off nucleons, as $\Tilde{\chi}_1 N\rightarrow \Tilde{\chi}_1 N$. Such limits are naturally dependent on the local DM density. The Xenon1T collaboration assume an isothermal DM profile for the galactic DM energy and velocity distribution, which leads to some implicit dependency on that profile in our direct detection study. However, the primary astrophysical dependency is on the local DM energy density which we take to be 0.3 GeV/cm$^3$, which is in agreement with recent calculations based on stellar kinematics~\cite{Cautun_milky_2020,Nitschai_First-Gaia_2020}.

Higgsinos can interact with matter primarily via an interaction mediated by a $Z$ or Higgs boson. As shown in the Lagrangian in Eq.~(\ref{eq:model}), there is no direct coupling of two pure higgsinos to a Higgs or $Z$ boson to allow for a scattering process. However, the mixing of higgsinos and gauginos allows a higgsino-Higgs-gaugino vertex to create a coupling of two $\Tilde{\chi}_1$ particles with a Higgs, with a similar procedure for a $Z$ boson. This dependency on mixing is the reason, as discussed in detail below, such couplings tend to fall off as $M_2$ gets large. Couplings via a $Z$ or Higgs boson require slightly different search strategies, as the $Z$ boson imposes a spin-dependent (SD) coupling, while the Higgs interaction is spin-independent (SI). While it often has a higher cross section, SD scattering is much harder to detect in modern experiments. SD cross sections go as the spin of the target nucleus, but SI cross sections tend to scale up with the size of the nucleus due to coherent interactions across the nucleons, which leads to several orders of magnitude improvement for SI in xenon~\cite{jungman_supersymmetric_1996,mayet_review_2016} as in Xenon1T. The end result is that the two types of scattering have similar exclusion limits, with SI resulting in  slightly stronger limits~\cite{Belanger_Discriminating_2009,Cohen_Correlation_2010,cheung_prospects_2013}, and so we will turn our attention toward Higgs-mediated interactions.

The cross section for a tree-level, Higgs-mediated interaction between higgsinos and nucleons can be analytically calculated~\cite{jungman_supersymmetric_1996, hisano_direct_2011, hisano_direct_2013}. While going forward in this paper we will use numerical methods due to issues arising at low $|\mu|$ and $M_2-|\mu|$, the analytic derivation is illustrated in Appendix~\ref{sec:AppendixDDscatter} for large $(M_2-|\mu|)/m_W$. The approximate result is
\begin{align}
   \sigma_{\rm SI} & \simeq 2.7\times 10^{-43}\pr{\frac{79.9 \GeV}{M_2-|\mu|} + \frac{24.1 \GeV}{M_1 - |\mu|}}^2 A~{\rm cm}^2 \nonumber \\
   & \simeq 1.6 \times 10^{-47} \left( \frac{A}{131} \right) \pr{\frac{10 \TeV}{M_2}}^2 {\rm cm}^2 , \label{eq:xsec}
\end{align}
for atomic mass $A$  of the target material (131 for xenon), and with the last form assuming $M_2 = M_1/3 \gg |\mu|$. As a reference, Xenon1T~\cite{aprile_dark_2018}, for DM mass $m_{\rm DM}$ well above 100 GeV, sets a limit following a rough power law
\begin{equation}
    \sigma_{\rm Xenon1T} \simeq 8\times 10^{-46} \pr{\frac{m_{\rm DM}}{\rm TeV}}~{\rm cm}^2 , \label{eq:exclusion}
\end{equation}
with the numerically calculated limit being roughly $M_2 \gtrsim 4$ TeV, regardless of the value of $\mu$.

\begin{figure}
    \centering
    \includegraphics[width=\imwidth]{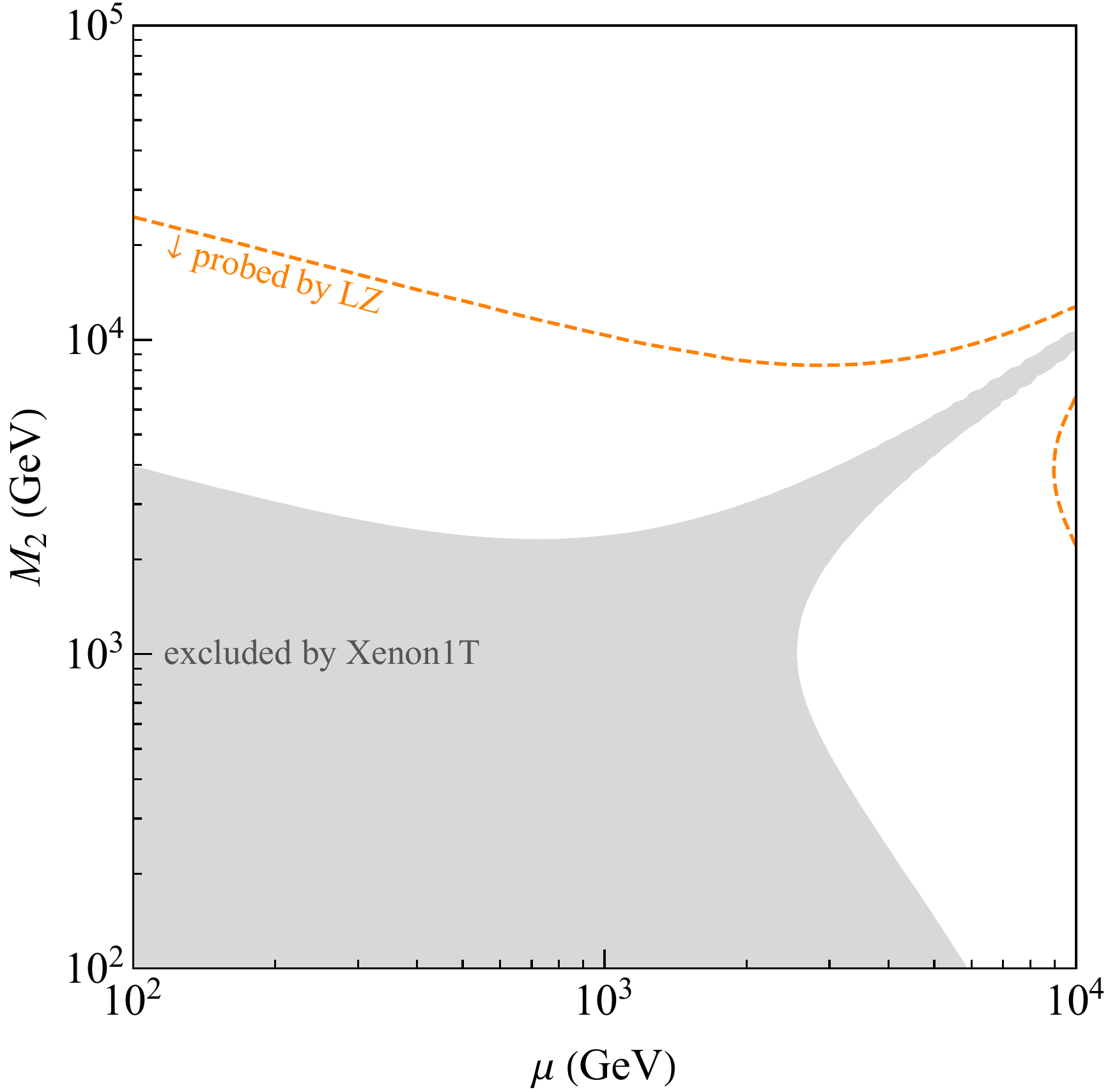}
    \caption{Excluded region for Xenon1T~\cite{xenon_1T_original} and projected reach for LZ~\cite{LZ_projected_2020} in searches for higgsino or wino DM with $\tan\beta = 10$. The gray shaded region is excluded, while the region below the dashed orange curve is expected to be within reach for LZ. The region above the $M_2=\mu$ diagonal represents higgsino DM.}
    \label{fig:direct_detection}
\end{figure}

Comparing the two above equations directly at $m_{\rm DM} \sim $ 1 TeV gives $M_2 \sim $ 1 TeV as the minimum allowed value from current bounds, which violates the condition that $M_2 \gg |\mu|$. It is therefore preferable to turn to computational tools for more precise cross section calculations. The code package \micromegas can calculate the cross section with nuclei for our model in the Lagrangian in Eq.~(\ref{eq:model}). Comparing the computational results to the constraints from Xenon1T yields the gray region in Fig.~\ref{fig:direct_detection}, and a similar comparison against projections for LUX-ZEPLIN (LZ)~\cite{LZ_projected_2020} yields the dashed orange curve in the same figure. For the latter case, we can see that the limits from LZ lie at high enough $M_2$ that Eq.~(\ref{eq:xsec}) fits the result well, with LZ limits following the same trend as in Eq.~(\ref{eq:exclusion}) for $m_{\rm DM} > 100 \GeV$, but with a factor of 40 smaller cross section.

It is worth noting that the limits from direct detection efforts are relatively weak for split higgsinos in the large $M_2$ limit. This is because the leading order nuclear scattering cross section is proportional to the degree of mixing between the higgsinos and gauginos. While the limits are robust for small mass splittings, as $M_2/\mu$ grows large, the LSP approaches a pure higgsino, and the nuclear cross section dramatically decreases, as shown in Eq.~(\ref{eq:xsec}).

\section{Indirect Detection}
\label{sec:ind_dec}

Indirect detection is a powerful tool for DM searches, in which telescope data are used to search for evidence of DM particles colliding and annihilating in regions of high DM density, such as the center of the Milky Way (for a review of the topic, see Ref.~\cite{slatyer_tasi_2017}). The upcoming Cherenkov Telescope Array (CTA)~\cite{CTA_Ground_2011} is expected to have significantly greater sensitivity than previous studies to such annihilation signals. This adds a powerful set of tools to DM detection, as the sensitivities depend on very different parameters than direct detection searches.

The most straightforward way to look for DM through indirect detection lies in the process $\Tilde{\chi}_1\Tilde{\chi}_1\rightarrow \gamma\gamma$. This creates a clean, monochromatic signal visible as a photon line in gamma-ray astronomy. For a given cross section times velocity for the annihilation process, $\sigma v$, we expect the gamma-ray flux for a photon line at energy $E$ through a patch of sky of solid angle $\Delta \Omega$ to be~\cite{hryczuk_testing_2019}
\begin{equation}
    \frac{\deriv\Phi_\gamma^{\rm DM}}{\deriv E}\pr{\Delta\Omega,E} = \frac{\sigma v}{8\pi m^2_{\rm DM}}\frac{\deriv N_\gamma\pr{E}}{\deriv E} J(\Delta \Omega) ,
\end{equation}
where $\frac{\deriv N_\gamma\pr{E}}{\deriv E}$ is the photon spectrum, and $J$ is a factor accounting for the DM density profile by integrating over the DM along a view in that solid angle. For the pure process $\Tilde{\chi}_1\Tilde{\chi}_1\rightarrow \gamma\gamma$, the flux integrates to just
\begin{equation}
    \Phi_\gamma^{\rm DM}\pr{\Delta\Omega} = \frac{\sigma v}{4\pi m^2_{\rm DM}} J(\Delta \Omega).
\end{equation}

The $J$-factor is~\cite{hryczuk_testing_2019}
\begin{equation}
    J(\Delta \Omega) \equiv \int_{\Delta \Omega} \deriv\Omega\int_0^\infty \deriv s~\rho_{\rm DM}^2 \pr{r(s,\theta_{\rm obs})}
\end{equation}
where $s$ is integrating along the line of sight from the telescope, and $\theta_{\rm obs}$ is the angle between that line and the galactic center, which together can be used to calculate the radius to the galactic center $r$, which in turn sets the DM density $\rho_{\rm DM}$. The density profile $\rho_{\rm DM}\pr{r}$ of DM is still a matter of significant study~\cite{de_blok_core-cusp_2010}, and 
in particular, there is debate between profiles with significant cusp-like behavior (sharply increasing density) near the galactic center, which are based in simulation, and observational evidence consistent with a core of constant density. For this reason, we use a range of $J$-factors, from ones based in the heavily cusped Einasto profile~\cite{Einasto:1965czb}
\begin{equation}
    \rho_{\rm Ein}\pr{r} = \rho_{0,{\rm Ein}} \exp\left\{ {-\frac{2}{a}\left[\pr{\frac{r}{r_s}}^a-1\right]}\right\} ,
\end{equation}
to the Navarro-Frenk-White profile~\cite{Navarro_Structure_1996},
\begin{equation}
    \rho_{\rm NFW}\pr{r}=\frac{\rho_{0,{\rm NFW}}}{\frac{r}{r_s}\pr{1+ \frac{r}{r_s}}^2} \ ,
\end{equation}
and further to a cored Einasto profile, in which the region out to 3 kpc from the galactic center is set to $\rho_{\rm DM} = \rho$(3 kpc), and outside 3 kpc it simply follows the Einasto curve~\cite{hryczuk_testing_2019}.

In these profiles, $r_s$ is the scale radius (roughly 20 kpc for the Milky Way~\cite{Pieri_Implications_2011}), $a$ is a factor tuning the slope of the cusp (roughly 0.17 for the Milky Way~\cite{Abramowski_Search_2011, slatyer_prospects_2021}), and the $\rho_0$'s are normalization factors tuned to account for the local DM density (roughly 0.3 GeV/cm$^3$~\cite{Cautun_milky_2020,Nitschai_First-Gaia_2020}).

\begin{figure}
    \centering
    \includegraphics[width=\imwidth]{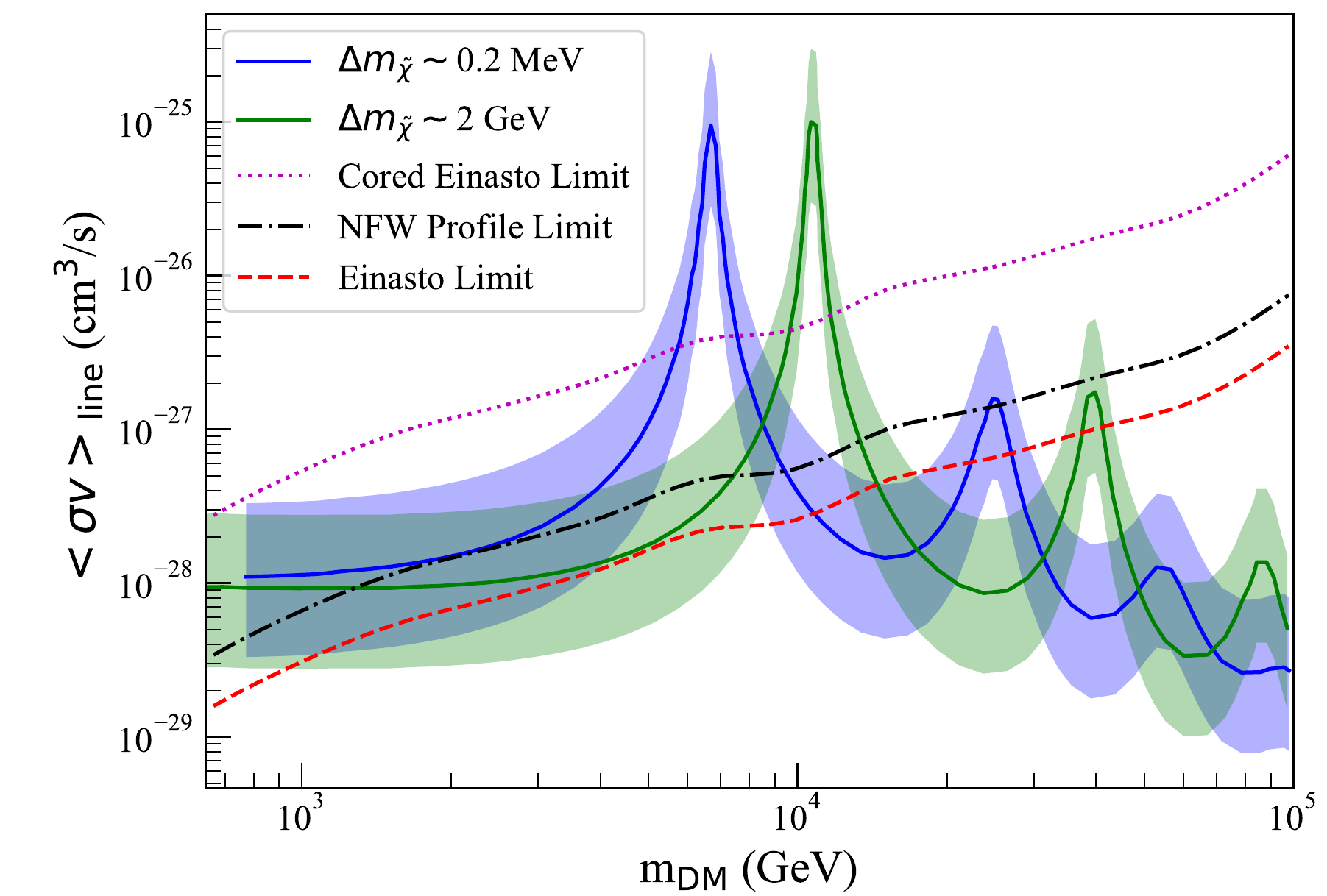}
    \caption{Expected higgsino DM discovery potential through searches for line photons in CTA, extracted from Refs.~\cite{hryczuk_testing_2019, slatyer_prospects_2021}. The limits are placed on expected annihilation cross section times velocity for higgsinos annihilating into photons, giving a monochromatic ``line'' signal. The green line is the theoretical prediction for the higgsino annihilation cross section assuming 2 GeV mass splitting, while the blue line is a phenomenological check at near-degenerate higgsinos, both extracted from Ref.~\cite{slatyer_prospects_2021}. The bands about each are factors of three, representing $\mathcal{O}(1)$ corrections in theory uncertainty. The expected reach for CTA assuming an Einasto DM density profile (red dashed) is extracted from Ref.~\cite{slatyer_prospects_2021}; the limit is offset (black dash-dotted and purple dotted) based on projections from Ref.~\cite{hryczuk_testing_2019} to show the reach due to the (NFW and cored Einasto) DM density profile assumptions. A point in theory space is detectable by CTA if the predicted line (blue or green) is above the limit curve (purple, black, or red).}
    \label{fig:line_photons}
\end{figure}

$\gamma$-ray telescope searches for line photons from DM annihilation can be particularly powerful in split higgsino searches as the annihilation process depends significantly less on the particulars of the model than other approaches, depending almost exclusively on the higgsino mass. On the other hand, such searches can be more dependent on different uncertainties, such as the DM profile, as described above, or theory corrections. Theory corrections arise primarily~\cite{slatyer_prospects_2021} due to Sommerfeld enhancement, low energy continuum emission from cascade decays, Sudakov double logarithm resummation, and inclusion of endpoint photons at energies near the peak. The Sommerfeld enhancements, due to corrections from electroweak particle exchanges at large distances~\cite{Hisano:2003ec,Hisano:2004ds}, have been calculated in Ref.~\cite{slatyer_prospects_2021}, but the remaining corrections are expected to yield $\mathcal{O}(1)$ corrections to the prediction.  Fig.~\ref{fig:line_photons} shows possible theory predictions extracted from Ref.~\cite{slatyer_prospects_2021} as solid lines, with the bands a factor of three about the prediction curve as an illustration of the uncertain corrections. As is visible in these curves, the Sommerfeld enhancement induces a peak-like structure to the theoretical prediction.

The non-solid lines in Fig.~\ref{fig:line_photons} show the expected reach for CTA, depending on the $J$-factor. The relative $J$-factors were extracted from Ref.~\cite{hryczuk_testing_2019} and used to rescale the projected discovery reach calculated in Ref.~\cite{slatyer_prospects_2021}. Of particular note, adjusting the $J$-factor leads to $\mathcal{O}(10)$ changes in discovery potential.

The discovery potential shown in Fig.~\ref{fig:line_photons} ranges from profound, reaching any higgsino of mass below 10 TeV, along with a band about 40 TeV, to the conservative, sensitive solely to bands centered between 6 and 10 TeV, depending on the mass splitting.

Annihilation to massive gauge bosons makes up a large fraction of the higgsino total cross section. It is large enough that despite the difficulty detecting gauge bosons through the photons created in their decays as compared to a photon line signal, such searches form a promising detection avenue. Unlike the line photon signal discussed above, corrections like Sommerfeld enhancement are expected to be relatively small for this channel~\cite{kowalska_discreet_2018}, so we use \micromegas to calculate the annihilation cross sections. The discovery channel is dominated by annihilation into $W$ bosons. For illustrative purposes, the result from \micromegas for $300 \GeV< |\mu| < M_2$ follows to within a few percent as
\begin{equation}
    \langle\sigma v\rangle_{\Tilde{\chi}_1\Tilde{\chi}_1\rightarrow \gamma X} \simeq 5.1\times 10^{-27}\pr{\frac{\rm TeV}{\mu}}^2 \text{cm}^3/\text{s} ,
\end{equation}
where $X$ is some particle(s) released either from the annihilation process directly, or a shower from the prompt particles. This can range from annihilation into $Z$ bosons which decay into showers and radiate off some number of photons, to direct annihilation into a pair of photons. For the Einasto DM profile, the reach for CTA is projected to be roughly $6\times 10^{-27}$ cm$^3$/s, which gives a maximum $m_{\rm DM}$ of 900 GeV.

We compared our calculated cross sections to projections for the reach for CTA~\cite{hryczuk_testing_2019}. Our results agreed with analytical calculations~\cite{krall_last_2018} (though they are slightly more conservative than the results in Ref.~\cite{arkani-hamed_well-tempered_2006}). The most sensitive limits placed by previous data arise from Planck measurements of the cosmic microwave background (CMB). These limits arise because annihilating DM on cosmological scales can cause notable perturbations to the CMB power spectrum. However, our calculations find that the limits from Planck data do not place limits on higgsino-like DM in the region of interest. Further, the maximum sensitivity possible due to cosmic limits is only expected to push the limit on higgsino-like DM to masses of a few hundred GeV~\cite{galli_cmb_2009,madhavacheril_current_2014}, and so limits from CMB measurements are left out of Fig.~\ref{fig:collected_lims}.

As shown by the purple dotted and dashed lines in Fig.~\ref{fig:collected_lims}, CTA data is expected to have significant discovery potential for higgsinos lighter than $\sim 1$ TeV. This result is independent of $M_2$ at large $M_2/\mu$, providing an exciting detection avenue covering a large region of parameter space. This is again subject to astrophysical uncertainties such as the DM profile, whereby the most pessimistic profile leaves higgsino undetectable through this channel.

\section{Electron EDM}
\label{sec:edm}

Experiments measuring the electron electric dipole moment, unlike the other approaches discussed above, do not rely on the probed new physics composing the entirety of the dark matter, nor on the cosmologies involved. Rather they focus on how the new physics changes the $ee\gamma$ vertex in a charge-parity symmetry violating way.

In the current leading experiment, Advanced Cold Molecular Electron EDM (ACME) II~\cite{ACME_improved_2018}, the \eEDM is measured using one of the strongest electric fields yet seen, the field inside a thorium monoxide molecule (ThO). The experiment focuses on an electron in the H$3\Delta_1$ electronic state, with spin pumped by lasers to a plane perpendicular to the field through a coherent superposition of the up and down spin eigenstates. The electron precesses as the molecule drifts through a chamber for a known amount of time, and the precession rate is found by measuring the final spin direction through fluorescence from laser excitation. A series of such molecules are aligned or anti-aligned in the field and give different precession frequencies that differ solely by the product of the dipole moment and the internal electric field. Using a series of sophisticated techniques to cool the molecules and shield them from ambient electromagnetic fields, along with reducing systematics using generated electric and magnetic fields, ACME II measured the \eEDM to be consistent with zero, with $d_e < 1.1\times 10^{-29}$ $e$ cm to 95\% confidence~\cite{ACME_improved_2018}. For comparison, the dipole moment for the electron in the Standard Model is expected to be at most $\mathcal{O}(10^{-38}) ~e$ cm~\cite{Pospelov_CKM_2014}.

\begin{figure}
    \centering
\includegraphics[width=0.5\linewidth]{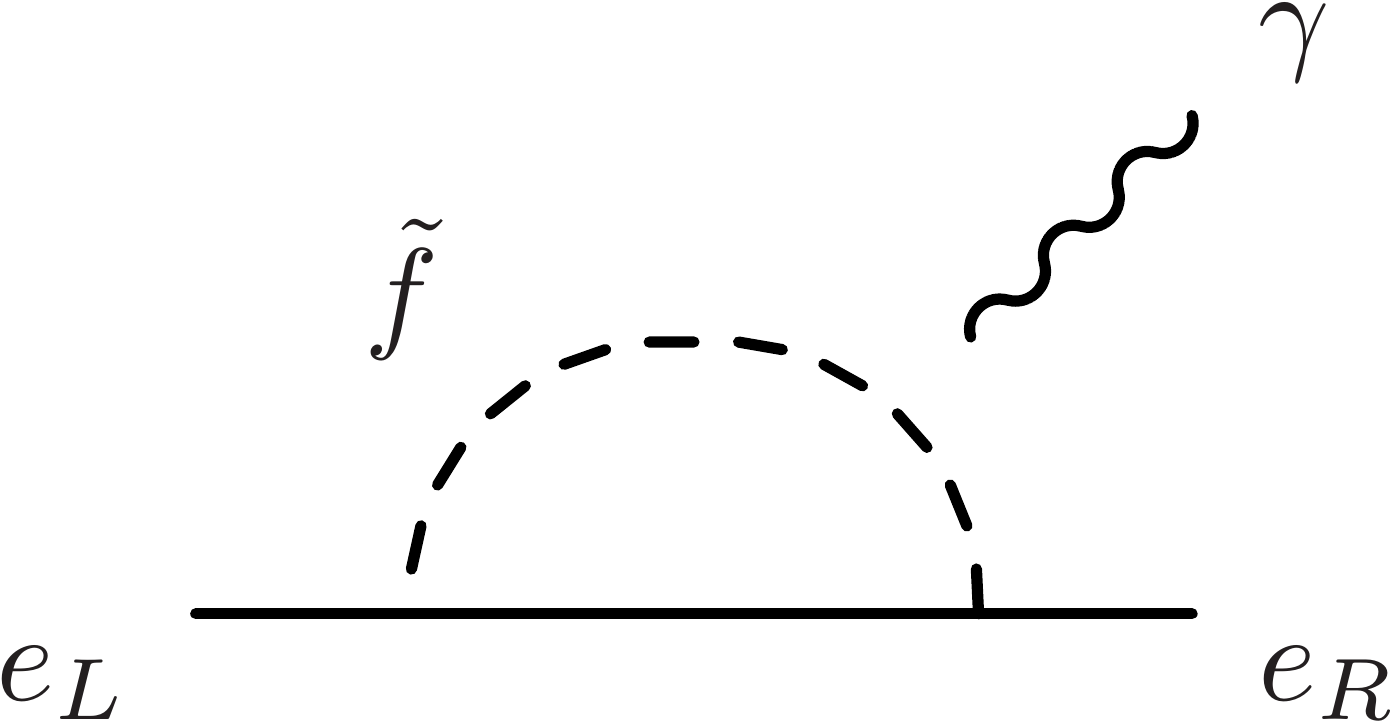}
    \caption{One-loop contributions to EDMs from SUSY. The loop is composed of a sfermion, $\Tilde{f}$, that is either a selectron or a sneutrino, with a neutralino or chargino (respectively) completing the loop and the external photon line to be attached to the charged internal line. These tend to be strongly suppressed for scenarios with heavy scalars.}
    \label{fig:1loop}
\end{figure}

Phases introduced in SUSY breaking terms can cause a large enough \eEDM at one-loop order to violate current measurements in models, like the MSSM, that have low scalar masses~\cite{abel_edm_2001,ibrahim_cp_2008}. This is achieved through diagrams shown in Fig.~\ref{fig:1loop} in the simplest SUSY models. This issue is sidestepped in the class of models discussed here by large suppression due to high sfermion masses, usually required by \eEDM limits to be $\gtrsim \mathcal{O}(10)$ TeV~\cite{cesarotti_interpreting_2019}.

\begin{figure}
    \centering
\includegraphics[width=0.3\linewidth]{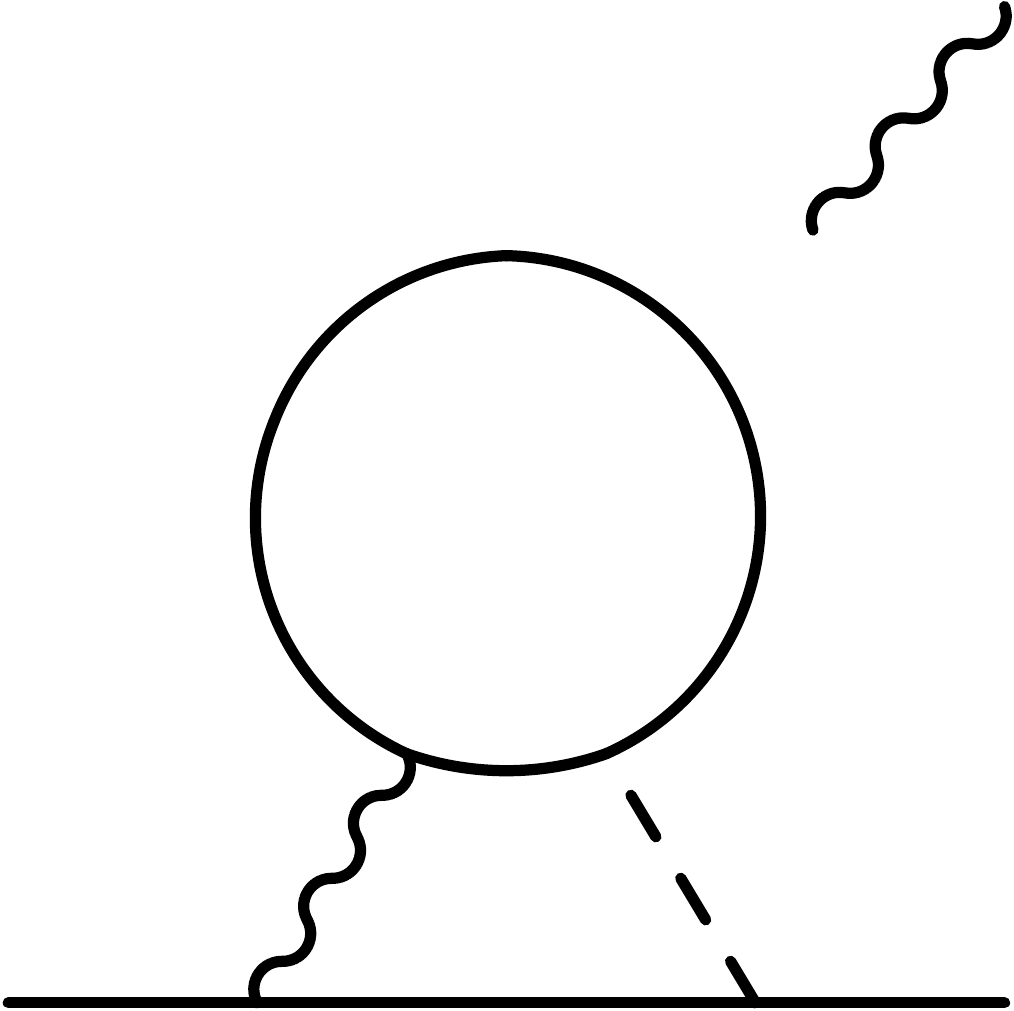}
\hspace{0.5 in}
\includegraphics[width=0.3\linewidth]{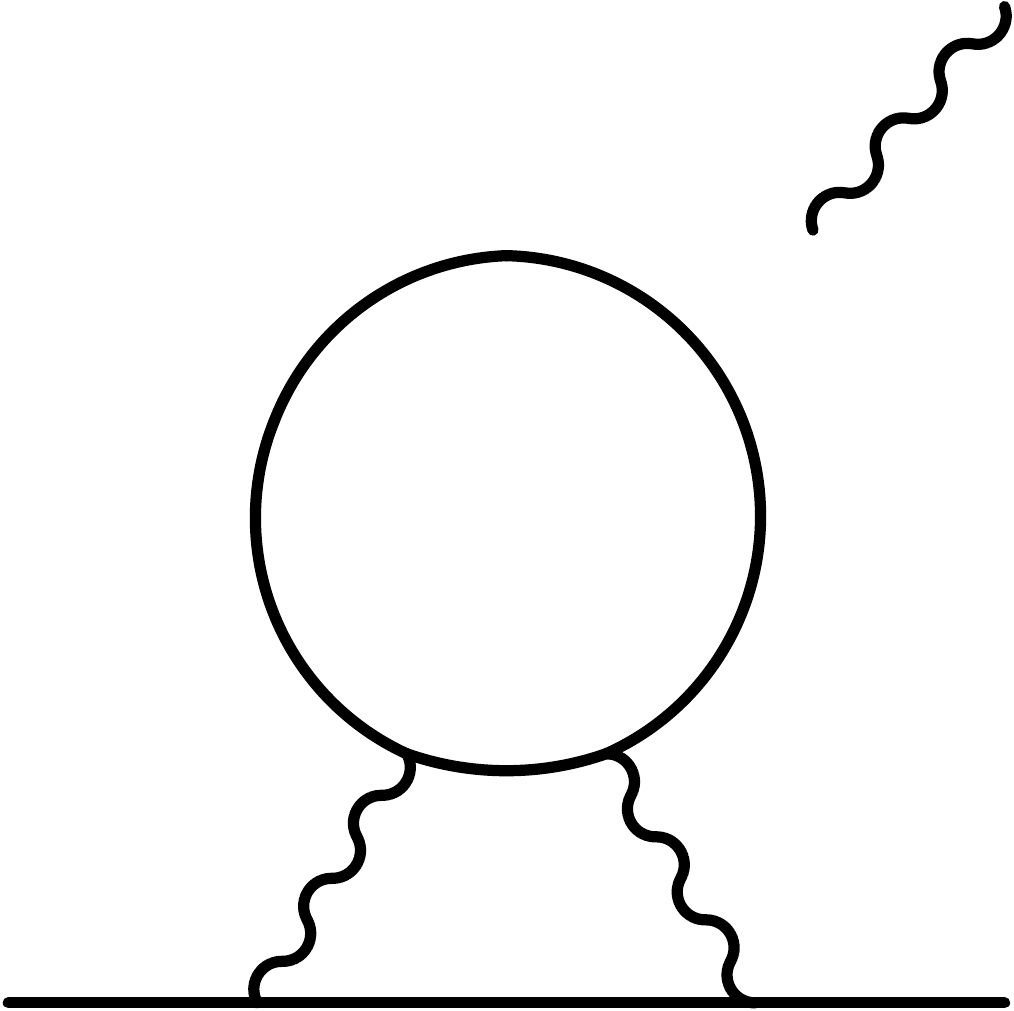}
    \caption{Barr-Zee diagrams~\cite{Barr_zee_edm} used to calculate the leading contribution to \eEDM for split higgsinos, where the external photon line is to be attached to any charged internal lines. The left diagram is a $Z$ boson or a photon and a Higgs connecting the electron line to a chargino loop, while the right diagram is two $W$-bosons connecting to a chargino-neutralino loop. The case of the right diagram with two $Z$-bosons has been found to be identically zero~\cite{giudice_electric_2006}.}
    \label{fig:2loop}
\end{figure}

Split higgsinos are not immune to EDM measurements, however. Chargino loops can induce an EDM at the two-loop level which can still be large enough for modern experiments to detect. There exist two classes of EDM contributions based on whether heavy Higgs bosons $H^\pm, H^0, A^0$ are involved. The contributions from $H^\pm, H^0, A^0$ are suppressed in the limit of heavy Higgs bosons. We focus on the case where only Standard Model Higgs and gauge bosons are responsible for mediating the two-loop diagrams and comment on the regions of the parameter space where heavy Higgs bosons become relevant. The contributions to the electron EDM $d_e$ mediated by $h, Z,$ and $W$ are separable into 3 components
\begin{equation}
\label{eq:edm_hZWg}
    d_e = d_{\gamma h} + d_{Z h} + d_{WW} ,
\end{equation}
where each $d_{ij}$ corresponds to the contribution to the \eEDM arising from the respective diagram in Fig.~\ref{fig:2loop}. In most cases within our region of interest, the hierarchy between the resultant contributions to \eEDM are $d_{\gamma h} > d_{Zh} > |d_{WW}|$ ($d_{WW}$ is negative in this region). The full expression is derived in detail in Ref.~\cite{giudice_electric_2006}, with the scaling evident in the limiting case, for $M_2, \mu \gg m_Z$,
\begin{align}
        d_{\gamma h} \simeq & \ \frac{-e\alpha m_e}{8\pi^3}\frac{\Tilde{g}_u\Tilde{g}_d}{M_2\mu} \sin{\phi_2} F_{\gamma h}\pr{\frac{M_2^2}{\mu^2},\frac{M_2\mu}{m_h^2}} \\
        d_{Zh} \simeq & \ \frac{e\pr{4\sin^2\theta_W-1}\alpha m_e}{32\pi^3\cos^2\theta_W}\frac{\Tilde{g}_u\Tilde{g}_d}{M_2\mu}\sin\phi_2 \\ & \times F_{Zh}\pr{\frac{m_Z^2}{m_h^2}, \frac{M_2^2}{\mu^2}, \frac{M_2\mu}{m_h^2}} \nonumber \\
        d_{WW} \simeq & \ \frac{-e\alpha m_e}{32\pi^3\sin^2\theta_W} \left[\frac{\Tilde{g}_u\Tilde{g}_d}{M_2\mu}\sin\phi_2 F_{WW}^{(2)}\pr{\frac{M_2^2}{\mu^2},\frac{M_2\mu}{m_h^2}} \right. \nonumber \\
        & \left. + \frac{\Tilde{g}_u'\Tilde{g}_d'}{M_1\mu}\sin\phi_1 F_{WW}^{(1)}\pr{\frac{M_1^2}{\mu^2},\frac{M_1\mu}{m_h^2}}\right] ,
\end{align}
where the Standard Model parameters, $e, m_e, \alpha,$ and $\theta_W$ are the electron charge and mass, the fine structure constant, and the weak mixing angle, respectively. The angles $\phi_i$ are the CP-violating complex phases on $M_1$ and $M_2$, and all other possible complex phases involved can be shifted through field redefinitions back to solely those two. We take the phases to be equal, and call them $\phi$, though they need not be in full generality.

The functions $F_{\gamma h}, F_{Zh}, F_{WW}^{(i)}$ are defined explicitly in Ref.~\cite{giudice_electric_2006}. For the sake of intuition for $|\mu| < M_2 \lesssim \mathcal{O}$(PeV), $F_{WW}^{(i)}$ lie between 0.5 and 2, and the $d_{WW}$ contribution is subdominant everywhere in the region of interest. The former two functions may be approximated to within 20\% within the region of interest as
\begin{align}
    F_{\gamma h}\pr{\frac{M_2^2}{\mu^2},\frac{M_2\mu}{m_h^2}} & \simeq 3.4\pr{\frac{|\mu|}{\rm TeV}}^{0.18} \\
    F_{Zh}\pr{\frac{m_Z^2}{m_h^2}, \frac{M_2^2}{\mu^2}, \frac{M_2\mu}{m_h^2}} &\simeq 1.2\pr{\frac{|\mu|}{\rm TeV}}^{0.3}\pr{\frac{\rm TeV}{M_2}}^{0.1} . \nonumber
\end{align} 
In either case, these functions have relatively small effects on scaling laws over model parameters.

The phase $\phi$ was taken to be $\pi/2$ for maximal CP violation and thereby a maximal \eEDM value. As the scaling goes as $\sin\phi$, the \eEDM does not drastically change while $\phi\sim \mathcal{O}$(1). The \eEDM calculation results were also verified against those attained using a similar model to ours in Ref.~\cite{cesarotti_interpreting_2019}. 

\begin{figure}
    \centering
    \includegraphics[width=\linewidth]{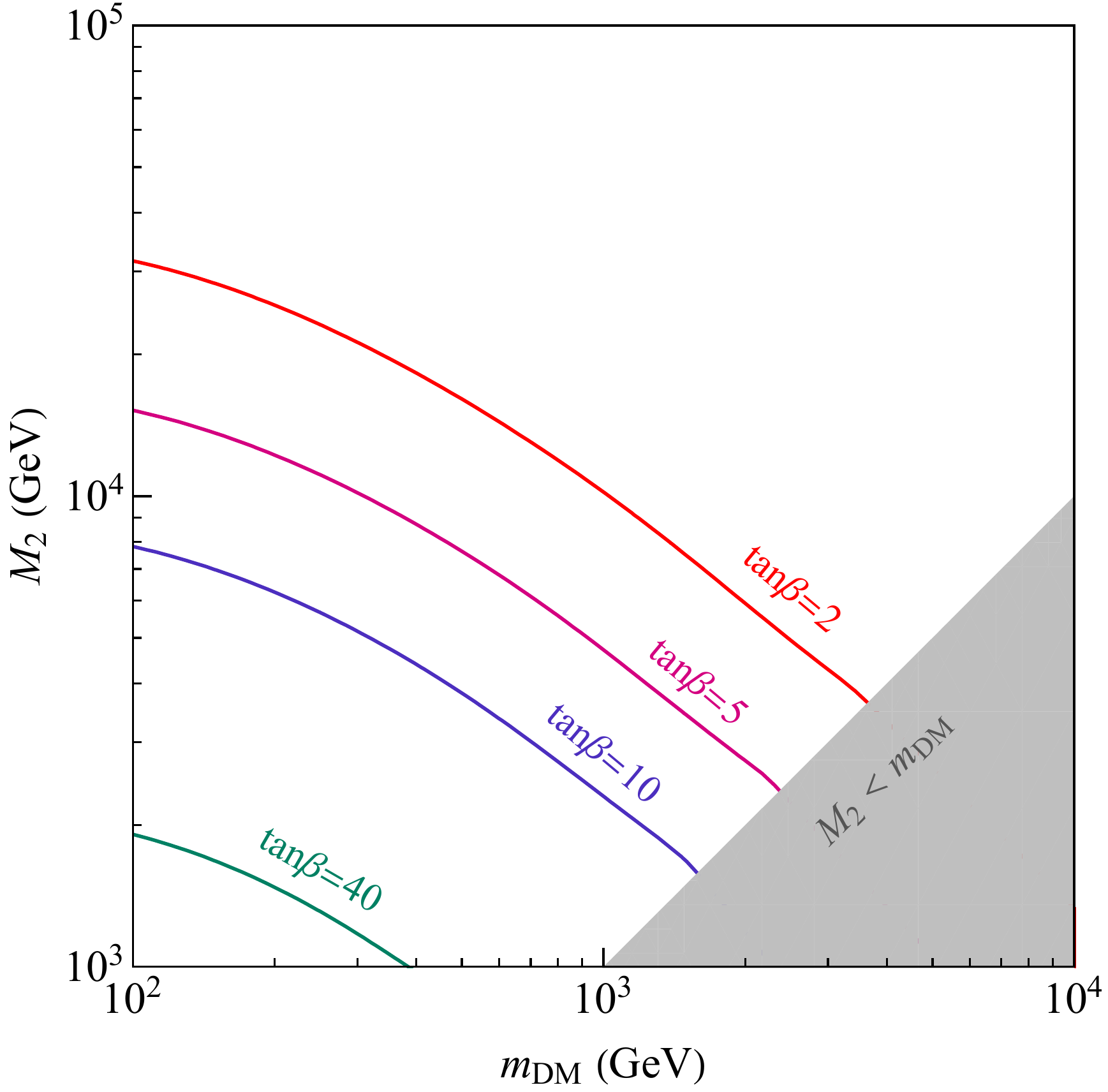}
    \caption{Exclusion limits from the \eEDM limit, $d_e < 1.1 \times 10^{-29} e~{\rm cm}$, obtained by ACME-II~\cite{ACME_improved_2018}. $M_2$ is assumed to be 0.3$M_1$, and $\phi = \pi/2$. The region very close to the diagonal is removed due to complications from the neutralino mass degeneracy, and below the diagonal is ignored as the wino becomes the LSP.}
    \label{fig:edm_lims}
\end{figure}

There are a few scaling laws of interest for the diagrams in Fig.~\ref{fig:2loop}. First is that the \eEDM scales as $\sin\beta\cos\beta$, due to flavor changing vertices in the gaugino loops. Contrary to what one usually finds for the one-loop \eEDM in SUSY, the two-loop dipole moment falls off for large $\tan\beta$. In addition, the electric dipole moment, $d_e$, falls both as $\mu$ and $M_2$, as increasing their masses suppresses the chargino loop. Lastly, the \eEDM varies directly as $\sin\phi$, which is expected as $\phi$ is the only CP-breaking phase involved.

The results of these calculations for the current sensitivity of ACME II~\cite{ACME_improved_2018} are shown as the solid lines in Fig.~\ref{fig:edm_lims} for different values of $\tan\beta$. Juxtaposed against other existing limits, both the current ACME II data (blue shaded region) and next generation Advanced ACME projections (blue dashed line) are shown in Fig.~\ref{fig:collected_lims}.

\begin{figure}
    \centering
    \includegraphics[width=\imwidth]{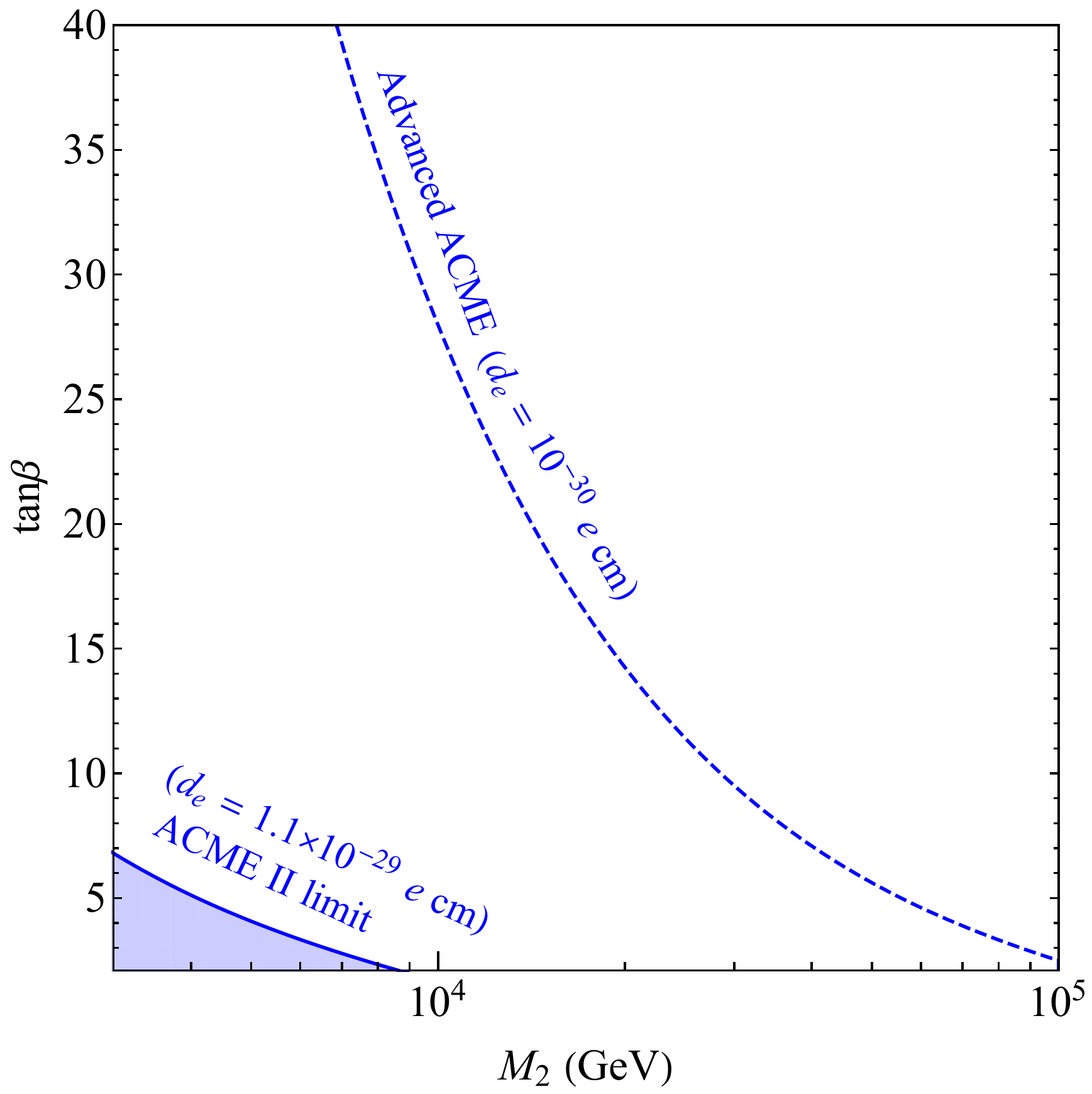}
    \caption{Sensitivities both from ACME-II~\cite{ACME_improved_2018} and projected next generation EDM limits~\cite{ACME_next_gen}, assuming $\mu = 1.2$ TeV, $M_2 = 0.3M_1$, and $\phi = \pi/2$. $M_2$ is restricted to be $> 3$ TeV to keep a relatively higgsino-like LSP.}
    \label{fig:tanbeta}
\end{figure}

Focusing just on the the region for the simple split higgsino model described in Sec.~\ref{sec:model}, with standard cosmology implying $m_{\rm DM} \simeq 1.2$ TeV, we find the current data only reaches very small $\tan\beta$. However, the next generation of \eEDM experiments are expected to be an order of magnitude more sensitive~\cite{ACME_next_gen}. We would then be able to detect models with quite large $\tan\beta$, as one can see comparing the curves in Fig.~\ref{fig:tanbeta}, where the blue solid (dashed) line is the current constraint (future prospect) for ACME. Even at a modest $\tan\beta = 10$, we can see in Fig.~\ref{fig:collected_lims} that the next generation of EDM experiments can have significant discovery potential. This makes a strong case for future \eEDM experiments as probes of new physics.

We now return to the \eEDM contributions involving the heavy Higgs bosons~\cite{Li:2008kz},
\begin{equation}
\label{eq:edm_HA}
    d_e = d_{\gamma H^0} + d_{\gamma A^0} + d_{W H^\pm} + d_{Z H^0} + d_{Z A^0} ,
\end{equation}
where each $d_{ij}$ denotes the contribution from the two-loop diagrams involving the corresponding heavy Higgs and gauge bosons. The full expressions are given in Ref.~\cite{Li:2008kz}. These contributions are suppressed by the masses of $H^\pm, H^0,$ and $A^0$ but enhanced for large $\tan \beta$. In particular, they are subdominant in our parameter space of interest to those in Eq.~(\ref{eq:edm_hZWg}) when $m_{A^0} \gtrsim (10, 100, 200, 1000) \TeV$ for $\tan\beta = (2, 5, 10, 40)$. We leave thorough exploration of the effects of these heavy Higgs bosons to future work.

\section{results}
\label{sec:results}

\begin{figure}
    \centering
    \includegraphics[width=\imwidth]{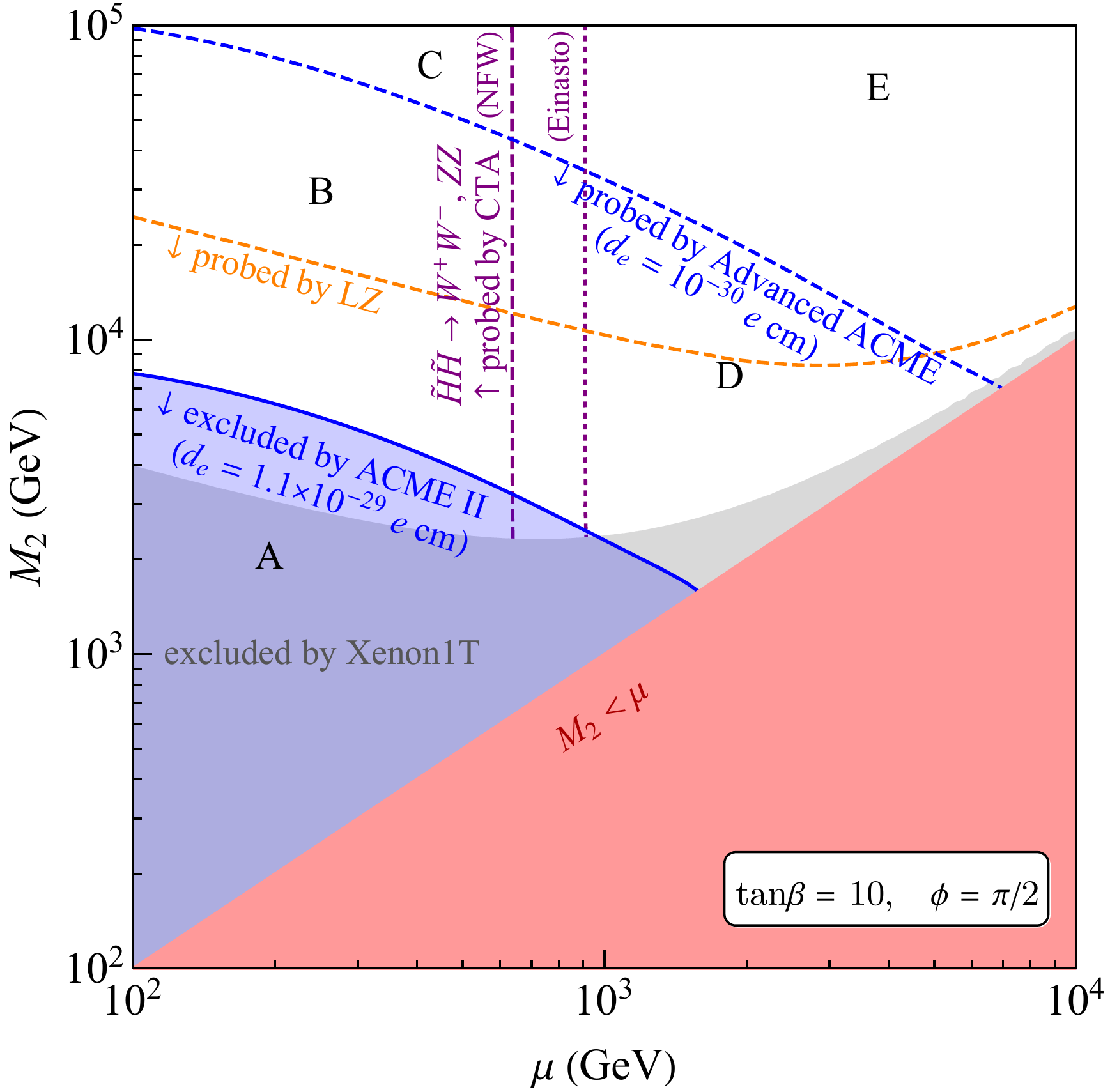}
    \caption{Collected limits from (gray region) direct detection by Xenon1T based on 2018 results, (orange dashes) projected reach for LZ~\cite{LZ_projected_2020}, (purple lines) indirect detection based on projected CTA results looking at annihilation to gauge bosons, particularly $W^+W^-$, with dotted and dashed reflecting Einasto and NFW DM profiles respectively, and (blue) from EDM constraints from ACME II data and projected future data, assuming split SUSY, higgsino LSP, $M_2 = 0.3M_1$, $\phi=\pi/2$, and tan$\beta = 10$. The capital letters indicate different regions where different search strategies dominate as discussed in Sec.~\ref{sec:results}}
    \label{fig:collected_lims}
\end{figure}

Accumulating the results of these three different techniques together, we get the various sensitivity curves and exclusion regions composing Fig.~\ref{fig:collected_lims}. The line photon results from indirect detection, discussed in Fig.~\ref{fig:line_photons}, are left out of this figure due to the large uncertainties involved. There are 5 different notable regions of the figure, each worth an individual mention, labelled with corresponding capital letters.
\begin{enumerate}[A.]
    \item This region is excluded by current experiments. For $M_2 \lesssim 4$ TeV as shown, the higgsino has a sufficient cross section with nucleons that we would expect to have seen scattering events at Xenon1T. Such low gaugino masses also lead to a large enough electron EDM to have been evident in ACME II data.
    \item This region is accessible to both indirect detection and \eEDM methods, and to a limited degree direct detection methods, in next generation experiments, such as CTA, Advanced ACME, and LZ respectively. In this regime we see the particular utility of the approaches in concert, due to their independent model dependencies. \eEDM measurements from Advanced ACME will be sensitive even for exotic DM density profiles or corrections to annihilation cross sections. Indirect detection results from CTA will remain viable even at large $\tan\beta$ and vanishing CP violation phase. Direct detection at LZ will be able to see higgsino DM even for small annihilation cross sections, and will depend only on the nucleon cross section and local DM density.
    \item This region can only be reached by future indirect detection experiments like CTA. Such techniques shine here, as the low energy effective theory for this model is simply the Standard Model with a new heavy, fermionic WIMP. The annihilation rate for the higgsino WIMP varies only with its mass, so such experiments can detect the higgsino for arbitrarily high gaugino masses.
    \item This region is accessible to future \eEDM experiments and direct detection, like Advanced ACME and LZ respectively. Again, these two methods complement each other, as the \eEDM does not depend on the likelihood for a DM particle to collide with another particle, while direct detection does not depend on complex phases and varies differently with model parameters. This area of parameter space is of particular interest because it neatly contains the preferred region for the most simple version of the split higgsino, outlined in Sec.~\ref{sec:model}, with standard cosmology creating higgsino DM through freeze-out, scalar masses all at $\mathcal{O}$(PeV), and gaugino masses generated through a conformal anomaly, which is all to say $M_2 \sim 10$ TeV, $\mu \sim 1.2$ TeV.
    \item This region is accessible only to monochromatic photon signatures in CTA. As discussed in Sec.~\ref{sec:ind_dec}, this is highly dependent on theory and DM profile unknowns. Nevertheless, it is likely that some areas in this region will be explored, particularly around $\mu \sim 10$ TeV where Sommerfeld enhancement implies a particularly large total annihilation cross section.
\end{enumerate}

\section{Conclusions}

We have analyzed split higgsinos over a wide possible parameter space, in particular focused on anomaly mediated SUSY breaking for the gaugino masses. Over a wide swath of this range, including the particularly favored model point (thermal relic DM candidate) of $M_2 = 0.3M_1 = 0.1M_3 =$ 3-10 TeV, with $|\mu| = 1.2$ TeV, $\sin\phi=1,$ and $\tan\beta=10$, we find previous experiments incapable of excluding the model, while next generation \eEDM experiments have substantial discovery potential.

The relative reach can be appreciated along a ray of $M_2 = 3|\mu|$, with the same relative values of $M_1$ and $M_3$ as above. For $|\mu| > 850$ GeV, neither ACME II nor Xenon1T can observe new physics effects, but LZ can observe a split higgsino along this ray for $|\mu| < 2.8$ TeV, and Advanced ACME can reach $|\mu| < 4$ TeV. Indirect detection is more variable in where exactly it can reach along this ray of model space, from the most optimistic dark matter density profiles implying CTA will have a discovery potential for $|\mu| \lesssim 10 \TeV$, and the most pessimistic profiles having little discovery potential much above the electroweak scale.

We see a real, significant discovery potential in the next generation experiments along these avenues, which is summarized by our Fig.~\ref{fig:collected_lims}. In particular,  Advanced ACME appears poised to present a robust, sizable and somewhat complementary reach in the split higgsino parameter space, though depending on various astrophysical parameters it still has a race against indirect and direct detection methods ahead of it. While much of this analysis focused on a particular realization involving anomaly mediated SUSY breaking, the conclusions tend to be fairly robust across the space of split higgsino models, depending primarily on higgsino masses, or on general neutralino mixing parameters that tend to depend on gaugino masses in ways similar to those in the construction discussed here.

{\bf Acknowledgments.}---
R.C.~would like to thank Keisuke Harigaya for useful discussions. The work was supported in part by the DoE DE-SC0007859 (J.W.), the DoE Early Career Grant DE-SC0019225 (R.C.), DE-SC0011842 at the University of Minnesota (R.C.), and NSF Graduate Research Fellowship (B.S.).

\appendix
\section{Direct Detection Scattering Cross Section}
\label{sec:AppendixDDscatter}

The scattering cross section for higgsinos scattering off a nucleon follows the formalism described in Ref.~\cite{jungman_supersymmetric_1996}. The full nuclear cross section from a Higgs mediated interaction is
\begin{equation}
    \sigma_{\rm SI} \approx \frac{4 m_r^2}{\pi} \pr{Zf_p + (A-Z)f_n}^2,
\end{equation}
where $A$ is the number of nucleons in the nucleus, $Z$ is the atomic number, and $m_r$ is the reduced mass $\frac{m_{\rm DM} m_N}{m_{\rm DM} + m_N}$ with $m_N$ the nucleon mass. $f_p/m_p$ and $f_n/m_n$ are the full tree level effective four-point vertex between 2 $\Tilde{\chi}_1$'s and 2 nucleons, 
\begin{align}
    \frac{f_p}{m_p} & \simeq \frac{f_n}{m_n} \simeq \pr{\sum_q f_q f_{T,q,n} - \pr{\frac{8\pi}{9}f_{T,G,n}f_G}} \\ 
    & \simeq \pr{0.06 f_q - 0.94 \alpha_S \frac{2}{9} f_q} \simeq -0.04f_q, \nonumber
\end{align}
where the $f_{T,q,n}$ and $f_{T,G,n}$ are nuclear form factors for quarks and gluons respectively. These form factors are for the neutron~\cite{jungman_supersymmetric_1996} and are approximately the same as for the proton (within 10\% after summation). $f_q$ and $f_G$ are effective four-point vertices between two neutralinos and two quarks or gluons respectively.

For $m_W \leq \mu, M_2, |M_2 - \mu|$, our tree level effective vertices with quarks are~\cite{hisano_direct_2013}
\begin{align}
\label{eq:quark}
    f_q = & \ -\frac{g^2}{4m_Wm_h^2}  \pr{Z_{\Tilde{\chi}_1,\Tilde{W}} - Z_{\Tilde{\chi}_1,\Tilde{B}} \tan\theta_W} \\ 
    & \times \pr{Z_{\Tilde{\chi}_1,\Tilde{H}_u}\cos\beta - Z_{\Tilde{\chi}_1,\Tilde{H}_d}\sin\beta} \nonumber \\
    \simeq & \ -\frac{g^2}{8m_h^2}  \pr{\frac{1}{|M_2| - |\mu|} + \frac{\tan^2\theta_W}{|M_1|-|\mu|}} \nonumber \\
    & \times \pr{1\pm\sin{2\beta}} 0.8 e^{0.2i}, \nonumber
\end{align}
where the mixing matrix $Z$ is defined such that $ZM_0Z^{\rm T} = M_D$, where $M_0$ is the neutralino mass matrix, and $M_D$ is the diagonal neutralino mass matrix. The form of the final equation is taken from Ref.~\cite{hisano_direct_2013}, with a factor of $0.8e^{0.2i}$ adjusting for the fact that the gaugino masses are assumed to have $\pi/2$ phases, and holds up well to within 10\%. The $\pm$ sign matches that of $\mu$. We take $\mu$ to be positive here, though the end result is the same cross section up to a factor of two. The value of $f_q$ does not depend on the quark due to flavor symmetry in the interaction. For gluons we have a similar calculation, as a result of the effective vertex arising from a heavy quark loop,
\begin{equation}
    f_G = \frac{\alpha_S}{12\pi}3f_q\label{eq:gluon}.
\end{equation}
The extra factor of 3 arises from a sum over charm, top, and bottom quarks.
Combining Eq.~(\ref{eq:quark}) and (\ref{eq:gluon}), this gives the cross section per nucleon of 
\begin{align}
   \sigma_{\rm SI} & \simeq  \frac{4 m_r^2}{\pi}\frac{\pr{Af_n}^2}{A} \\
   & \simeq 2.2\times 10^{-43} {\rm cm}^2 A \pr{\frac{79.9 \GeV}{|M_2|-|\mu|} + \frac{24.1 \GeV}{|M_1| - |\mu|}}^2  \nonumber \\
   & \simeq 1.3 \times 10^{-47} {\rm cm}^2 \left(\frac{A}{131}\right) \pr{\frac{10 \TeV}{M_2}}^2 , \nonumber
\end{align}
where the final expression assumes $M_1 = 3 M_2$ and $A$ is set to the atomic mass of xenon $A=131$.

Preliminary investigations indicate a moderate contribution at loop order~\cite{hisano_direct_2011,hisano_direct_2013, hill_wimp-nucleon_2014}, though the effects tend to be most pronounced for mixed neutralinos, i.e., for $M_2$ near $|\mu|$ and is not in the main interest of our analysis.

\bibliography{Higgsino_DM}

\begin{thebibliography}{68}%
\makeatletter
\providecommand \@ifxundefined [1]{%
 \@ifx{#1\undefined}
}%
\providecommand \@ifnum [1]{%
 \ifnum #1\expandafter \@firstoftwo
 \else \expandafter \@secondoftwo
 \fi
}%
\providecommand \@ifx [1]{%
 \ifx #1\expandafter \@firstoftwo
 \else \expandafter \@secondoftwo
 \fi
}%
\providecommand \natexlab [1]{#1}%
\providecommand \enquote  [1]{``#1''}%
\providecommand \bibnamefont  [1]{#1}%
\providecommand \bibfnamefont [1]{#1}%
\providecommand \citenamefont [1]{#1}%
\providecommand \href@noop [0]{\@secondoftwo}%
\providecommand \href [0]{\begingroup \@sanitize@url \@href}%
\providecommand \@href[1]{\@@startlink{#1}\@@href}%
\providecommand \@@href[1]{\endgroup#1\@@endlink}%
\providecommand \@sanitize@url [0]{\catcode `\\12\catcode `\$12\catcode
  `\&12\catcode `\#12\catcode `\^12\catcode `\_12\catcode `\%12\relax}%
\providecommand \@@startlink[1]{}%
\providecommand \@@endlink[0]{}%
\providecommand \url  [0]{\begingroup\@sanitize@url \@url }%
\providecommand \@url [1]{\endgroup\@href {#1}{\urlprefix }}%
\providecommand \urlprefix  [0]{URL }%
\providecommand \Eprint [0]{\href }%
\providecommand \doibase [0]{http://dx.doi.org/}%
\providecommand \selectlanguage [0]{\@gobble}%
\providecommand \bibinfo  [0]{\@secondoftwo}%
\providecommand \bibfield  [0]{\@secondoftwo}%
\providecommand \translation [1]{[#1]}%
\providecommand \BibitemOpen [0]{}%
\providecommand \bibitemStop [0]{}%
\providecommand \bibitemNoStop [0]{.\EOS\space}%
\providecommand \EOS [0]{\spacefactor3000\relax}%
\providecommand \BibitemShut  [1]{\csname bibitem#1\endcsname}%
\let\auto@bib@innerbib\@empty
\bibitem [{\citenamefont {{Zwicky}}(1933)}]{Zwicky_DM_1933}%
  \BibitemOpen
  \bibfield  {author} {\bibinfo {author} {\bibfnamefont {F.}~\bibnamefont
  {{Zwicky}}},\ }\href@noop {} {\bibfield  {journal} {\bibinfo  {journal}
  {Helvetica Physica Acta}\ }\textbf {\bibinfo {volume} {6}},\ \bibinfo {pages}
  {110} (\bibinfo {year} {1933})}\BibitemShut {NoStop}%
\bibitem [{\citenamefont {Kolb}\ and\ \citenamefont
  {Turner}(1988)}]{Kolb_Turner}%
  \BibitemOpen
  \bibfield  {author} {\bibinfo {author} {\bibfnamefont {E.~W.}\ \bibnamefont
  {Kolb}}\ and\ \bibinfo {author} {\bibfnamefont {M.~S.}\ \bibnamefont
  {Turner}},\ }\href@noop {} {\emph {\bibinfo {title} {{The Early Universe}}}}\
  (\bibinfo  {publisher} {Addison-Wesley Publishing Company},\ \bibinfo {year}
  {1988})\BibitemShut {NoStop}%
\bibitem [{\citenamefont {Alves}(2012)}]{Alves_Simplified_2012}%
  \BibitemOpen
  \bibfield  {author} {\bibinfo {author} {\bibfnamefont {D.}~\bibnamefont
  {Alves}} (\bibinfo {collaboration} {LHC New Physics Working Group}),\ }\href
  {\doibase 10.1088/0954-3899/39/10/105005} {\bibfield  {journal} {\bibinfo
  {journal} {J. Phys. G}\ }\textbf {\bibinfo {volume} {39}},\ \bibinfo {pages}
  {105005} (\bibinfo {year} {2012})},\ \Eprint {http://arxiv.org/abs/1105.2838}
  {arXiv:1105.2838 [hep-ph]} \BibitemShut {NoStop}%
\bibitem [{\citenamefont {Canepa}(2019)}]{Canepa_Searches_2019}%
  \BibitemOpen
  \bibfield  {author} {\bibinfo {author} {\bibfnamefont {A.}~\bibnamefont
  {Canepa}},\ }\href {\doibase 10.1016/j.revip.2019.100033} {\bibfield
  {journal} {\bibinfo  {journal} {Rev. Phys.}\ }\textbf {\bibinfo {volume}
  {4}},\ \bibinfo {pages} {100033} (\bibinfo {year} {2019})}\BibitemShut
  {NoStop}%
\bibitem [{\citenamefont {Baer}\ \emph {et~al.}(2020)\citenamefont {Baer},
  \citenamefont {Barger}, \citenamefont {Salam}, \citenamefont {Sengupta},\
  and\ \citenamefont {Sinha}}]{Baer:2020kwz}%
  \BibitemOpen
  \bibfield  {author} {\bibinfo {author} {\bibfnamefont {H.}~\bibnamefont
  {Baer}}, \bibinfo {author} {\bibfnamefont {V.}~\bibnamefont {Barger}},
  \bibinfo {author} {\bibfnamefont {S.}~\bibnamefont {Salam}}, \bibinfo
  {author} {\bibfnamefont {D.}~\bibnamefont {Sengupta}}, \ and\ \bibinfo
  {author} {\bibfnamefont {K.}~\bibnamefont {Sinha}},\ }\href {\doibase
  10.1140/epjst/e2020-000020-x} {\bibfield  {journal} {\bibinfo  {journal}
  {Eur. Phys. J. ST}\ }\textbf {\bibinfo {volume} {229}},\ \bibinfo {pages}
  {3085} (\bibinfo {year} {2020})},\ \Eprint {http://arxiv.org/abs/2002.03013}
  {arXiv:2002.03013 [hep-ph]} \BibitemShut {NoStop}%
\bibitem [{\citenamefont {Gaillard}\ and\ \citenamefont
  {Lee}(1974)}]{Gaillard_Rare_1974}%
  \BibitemOpen
  \bibfield  {author} {\bibinfo {author} {\bibfnamefont {M.~K.}\ \bibnamefont
  {Gaillard}}\ and\ \bibinfo {author} {\bibfnamefont {B.~W.}\ \bibnamefont
  {Lee}},\ }\href {\doibase 10.1103/PhysRevD.10.897} {\bibfield  {journal}
  {\bibinfo  {journal} {Phys. Rev. D}\ }\textbf {\bibinfo {volume} {10}},\
  \bibinfo {pages} {897} (\bibinfo {year} {1974})}\BibitemShut {NoStop}%
\bibitem [{\citenamefont {Ellis}\ and\ \citenamefont
  {Nanopoulos}(1982)}]{Ellis_Flavor_1981}%
  \BibitemOpen
  \bibfield  {author} {\bibinfo {author} {\bibfnamefont {J.~R.}\ \bibnamefont
  {Ellis}}\ and\ \bibinfo {author} {\bibfnamefont {D.~V.}\ \bibnamefont
  {Nanopoulos}},\ }\href {\doibase 10.1016/0370-2693(82)90948-0} {\bibfield
  {journal} {\bibinfo  {journal} {Phys. Lett. B}\ }\textbf {\bibinfo {volume}
  {110}},\ \bibinfo {pages} {44} (\bibinfo {year} {1982})}\BibitemShut
  {NoStop}%
\bibitem [{\citenamefont {Donoghue}\ \emph {et~al.}(1983)\citenamefont
  {Donoghue}, \citenamefont {Nilles},\ and\ \citenamefont
  {Wyler}}]{Donoghue_Flavor_1983}%
  \BibitemOpen
  \bibfield  {author} {\bibinfo {author} {\bibfnamefont {J.~F.}\ \bibnamefont
  {Donoghue}}, \bibinfo {author} {\bibfnamefont {H.~P.}\ \bibnamefont
  {Nilles}}, \ and\ \bibinfo {author} {\bibfnamefont {D.}~\bibnamefont
  {Wyler}},\ }\href {\doibase 10.1016/0370-2693(83)90072-2} {\bibfield
  {journal} {\bibinfo  {journal} {Phys. Lett. B}\ }\textbf {\bibinfo {volume}
  {128}},\ \bibinfo {pages} {55} (\bibinfo {year} {1983})}\BibitemShut
  {NoStop}%
\bibitem [{\citenamefont {Hall}\ \emph {et~al.}(1986)\citenamefont {Hall},
  \citenamefont {Kostelecky},\ and\ \citenamefont {Raby}}]{Hall:1985dx}%
  \BibitemOpen
  \bibfield  {author} {\bibinfo {author} {\bibfnamefont {L.~J.}\ \bibnamefont
  {Hall}}, \bibinfo {author} {\bibfnamefont {V.~A.}\ \bibnamefont
  {Kostelecky}}, \ and\ \bibinfo {author} {\bibfnamefont {S.}~\bibnamefont
  {Raby}},\ }\href {\doibase 10.1016/0550-3213(86)90397-4} {\bibfield
  {journal} {\bibinfo  {journal} {Nucl. Phys. B}\ }\textbf {\bibinfo {volume}
  {267}},\ \bibinfo {pages} {415} (\bibinfo {year} {1986})}\BibitemShut
  {NoStop}%
\bibitem [{\citenamefont {Baer}\ \emph {et~al.}(2019)\citenamefont {Baer},
  \citenamefont {Barger},\ and\ \citenamefont
  {Sengupta}}]{Baer_Landscape_2019}%
  \BibitemOpen
  \bibfield  {author} {\bibinfo {author} {\bibfnamefont {H.}~\bibnamefont
  {Baer}}, \bibinfo {author} {\bibfnamefont {V.}~\bibnamefont {Barger}}, \ and\
  \bibinfo {author} {\bibfnamefont {D.}~\bibnamefont {Sengupta}},\ }\href
  {\doibase 10.1103/PhysRevResearch.1.033179} {\bibfield  {journal} {\bibinfo
  {journal} {Phys. Rev. Res.}\ }\textbf {\bibinfo {volume} {1}},\ \bibinfo
  {pages} {033179} (\bibinfo {year} {2019})},\ \Eprint
  {http://arxiv.org/abs/1910.00090} {arXiv:1910.00090 [hep-ph]} \BibitemShut
  {NoStop}%
\bibitem [{\citenamefont {Gabbiani}\ \emph {et~al.}(1996)\citenamefont
  {Gabbiani}, \citenamefont {Gabrielli}, \citenamefont {Masiero},\ and\
  \citenamefont {Silvestrini}}]{Gabbiani:1996hi}%
  \BibitemOpen
  \bibfield  {author} {\bibinfo {author} {\bibfnamefont {F.}~\bibnamefont
  {Gabbiani}}, \bibinfo {author} {\bibfnamefont {E.}~\bibnamefont {Gabrielli}},
  \bibinfo {author} {\bibfnamefont {A.}~\bibnamefont {Masiero}}, \ and\
  \bibinfo {author} {\bibfnamefont {L.}~\bibnamefont {Silvestrini}},\ }\href
  {\doibase 10.1016/0550-3213(96)00390-2} {\bibfield  {journal} {\bibinfo
  {journal} {Nucl. Phys. B}\ }\textbf {\bibinfo {volume} {477}},\ \bibinfo
  {pages} {321} (\bibinfo {year} {1996})},\ \Eprint
  {http://arxiv.org/abs/hep-ph/9604387} {arXiv:hep-ph/9604387} \BibitemShut
  {NoStop}%
\bibitem [{\citenamefont {Weinberg}(1982)}]{Weinberg_Supersymmetry_1982}%
  \BibitemOpen
  \bibfield  {author} {\bibinfo {author} {\bibfnamefont {S.}~\bibnamefont
  {Weinberg}},\ }\href {\doibase 10.1103/PhysRevD.26.287} {\bibfield  {journal}
  {\bibinfo  {journal} {Phys. Rev. D}\ }\textbf {\bibinfo {volume} {26}},\
  \bibinfo {pages} {287} (\bibinfo {year} {1982})}\BibitemShut {NoStop}%
\bibitem [{\citenamefont {Sakai}\ and\ \citenamefont
  {Yanagida}(1982)}]{Sakai_Proton_1981}%
  \BibitemOpen
  \bibfield  {author} {\bibinfo {author} {\bibfnamefont {N.}~\bibnamefont
  {Sakai}}\ and\ \bibinfo {author} {\bibfnamefont {T.}~\bibnamefont
  {Yanagida}},\ }\href {\doibase 10.1016/0550-3213(82)90457-6} {\bibfield
  {journal} {\bibinfo  {journal} {Nucl. Phys. B}\ }\textbf {\bibinfo {volume}
  {197}},\ \bibinfo {pages} {533} (\bibinfo {year} {1982})}\BibitemShut
  {NoStop}%
\bibitem [{\citenamefont {Chamoun}\ \emph {et~al.}(2020)\citenamefont
  {Chamoun}, \citenamefont {Domingo},\ and\ \citenamefont
  {Dreiner}}]{Chamoun_Nucleon_2020}%
  \BibitemOpen
  \bibfield  {author} {\bibinfo {author} {\bibfnamefont {N.}~\bibnamefont
  {Chamoun}}, \bibinfo {author} {\bibfnamefont {F.}~\bibnamefont {Domingo}}, \
  and\ \bibinfo {author} {\bibfnamefont {H.~K.}\ \bibnamefont {Dreiner}},\
  }\href@noop {} {\  (\bibinfo {year} {2020})},\ \Eprint
  {http://arxiv.org/abs/2012.11623} {arXiv:2012.11623 [hep-ph]} \BibitemShut
  {NoStop}%
\bibitem [{\citenamefont {del Aguila}\ \emph {et~al.}(1983)\citenamefont {del
  Aguila}, \citenamefont {Gavela}, \citenamefont {Grifols},\ and\ \citenamefont
  {Mendez}}]{delAguila:1983dfr}%
  \BibitemOpen
  \bibfield  {author} {\bibinfo {author} {\bibfnamefont {F.}~\bibnamefont {del
  Aguila}}, \bibinfo {author} {\bibfnamefont {M.~B.}\ \bibnamefont {Gavela}},
  \bibinfo {author} {\bibfnamefont {J.~A.}\ \bibnamefont {Grifols}}, \ and\
  \bibinfo {author} {\bibfnamefont {A.}~\bibnamefont {Mendez}},\ }\href
  {\doibase 10.1016/0370-2693(83)90018-7} {\bibfield  {journal} {\bibinfo
  {journal} {Phys. Lett. B}\ }\textbf {\bibinfo {volume} {126}},\ \bibinfo
  {pages} {71} (\bibinfo {year} {1983})},\ \bibinfo {note} {[Erratum:
  Phys.Lett.B 129, 473 (1983)]}\BibitemShut {NoStop}%
\bibitem [{\citenamefont {Cesarotti}\ \emph {et~al.}(2019)\citenamefont
  {Cesarotti}, \citenamefont {Lu}, \citenamefont {Nakai}, \citenamefont
  {Parikh},\ and\ \citenamefont {Reece}}]{cesarotti_interpreting_2019}%
  \BibitemOpen
  \bibfield  {author} {\bibinfo {author} {\bibfnamefont {C.}~\bibnamefont
  {Cesarotti}}, \bibinfo {author} {\bibfnamefont {Q.}~\bibnamefont {Lu}},
  \bibinfo {author} {\bibfnamefont {Y.}~\bibnamefont {Nakai}}, \bibinfo
  {author} {\bibfnamefont {A.}~\bibnamefont {Parikh}}, \ and\ \bibinfo {author}
  {\bibfnamefont {M.}~\bibnamefont {Reece}},\ }\href {\doibase
  10.1007/JHEP05(2019)059} {\bibfield  {journal} {\bibinfo  {journal} {Journal
  of High Energy Physics}\ }\textbf {\bibinfo {volume} {2019}},\ \bibinfo
  {pages} {59} (\bibinfo {year} {2019})},\ \bibinfo {note} {arXiv:
  1810.07736}\BibitemShut {NoStop}%
\bibitem [{\citenamefont {Wells}(2003)}]{wells_implications_2003}%
  \BibitemOpen
  \bibfield  {author} {\bibinfo {author} {\bibfnamefont {J.~D.}\ \bibnamefont
  {Wells}},\ }\href@noop {} {\bibfield  {journal} {\bibinfo  {journal}
  {arXiv:hep-ph/0306127}\ } (\bibinfo {year} {2003})},\ \bibinfo {note} {arXiv:
  hep-ph/0306127}\BibitemShut {NoStop}%
\bibitem [{\citenamefont {Arkani-Hamed}\ \emph {et~al.}(2006)\citenamefont
  {Arkani-Hamed}, \citenamefont {Delgado},\ and\ \citenamefont
  {Giudice}}]{arkani-hamed_well-tempered_2006}%
  \BibitemOpen
  \bibfield  {author} {\bibinfo {author} {\bibfnamefont {N.}~\bibnamefont
  {Arkani-Hamed}}, \bibinfo {author} {\bibfnamefont {A.}~\bibnamefont
  {Delgado}}, \ and\ \bibinfo {author} {\bibfnamefont {G.~F.}\ \bibnamefont
  {Giudice}},\ }\href {\doibase 10.1016/j.nuclphysb.2006.02.010} {\bibfield
  {journal} {\bibinfo  {journal} {Nuclear Physics B}\ }\textbf {\bibinfo
  {volume} {741}},\ \bibinfo {pages} {108} (\bibinfo {year} {2006})},\ \bibinfo
  {note} {arXiv: hep-ph/0601041}\BibitemShut {NoStop}%
\bibitem [{\citenamefont {Baer}\ \emph {et~al.}(2016)\citenamefont {Baer},
  \citenamefont {Barger},\ and\ \citenamefont {Serce}}]{Baer:2016ucr}%
  \BibitemOpen
  \bibfield  {author} {\bibinfo {author} {\bibfnamefont {H.}~\bibnamefont
  {Baer}}, \bibinfo {author} {\bibfnamefont {V.}~\bibnamefont {Barger}}, \ and\
  \bibinfo {author} {\bibfnamefont {H.}~\bibnamefont {Serce}},\ }\href
  {\doibase 10.1103/PhysRevD.94.115019} {\bibfield  {journal} {\bibinfo
  {journal} {Phys. Rev. D}\ }\textbf {\bibinfo {volume} {94}},\ \bibinfo
  {pages} {115019} (\bibinfo {year} {2016})},\ \Eprint
  {http://arxiv.org/abs/1609.06735} {arXiv:1609.06735 [hep-ph]} \BibitemShut
  {NoStop}%
\bibitem [{\citenamefont {Kowalska}\ and\ \citenamefont
  {Sessolo}(2018)}]{kowalska_discreet_2018}%
  \BibitemOpen
  \bibfield  {author} {\bibinfo {author} {\bibfnamefont {K.}~\bibnamefont
  {Kowalska}}\ and\ \bibinfo {author} {\bibfnamefont {E.~M.}\ \bibnamefont
  {Sessolo}},\ }\href {\doibase 10.1155/2018/6828560} {\bibfield  {journal}
  {\bibinfo  {journal} {Advances in High Energy Physics}\ }\textbf {\bibinfo
  {volume} {2018}},\ \bibinfo {pages} {1} (\bibinfo {year} {2018})},\ \bibinfo
  {note} {arXiv: 1802.04097}\BibitemShut {NoStop}%
\bibitem [{\citenamefont {Profumo}\ and\ \citenamefont
  {Yaguna}(2004)}]{Profumo_Statistical_2004}%
  \BibitemOpen
  \bibfield  {author} {\bibinfo {author} {\bibfnamefont {S.}~\bibnamefont
  {Profumo}}\ and\ \bibinfo {author} {\bibfnamefont {C.~E.}\ \bibnamefont
  {Yaguna}},\ }\href {\doibase 10.1103/PhysRevD.70.095004} {\bibfield
  {journal} {\bibinfo  {journal} {Phys. Rev. D}\ }\textbf {\bibinfo {volume}
  {70}},\ \bibinfo {pages} {095004} (\bibinfo {year} {2004})},\ \Eprint
  {http://arxiv.org/abs/hep-ph/0407036} {arXiv:hep-ph/0407036} \BibitemShut
  {NoStop}%
\bibitem [{\citenamefont {Giudice}\ and\ \citenamefont
  {Romanino}(2005)}]{giudice_split_2005}%
  \BibitemOpen
  \bibfield  {author} {\bibinfo {author} {\bibfnamefont {G.~F.}\ \bibnamefont
  {Giudice}}\ and\ \bibinfo {author} {\bibfnamefont {A.}~\bibnamefont
  {Romanino}},\ }\href {\doibase 10.1016/j.nuclphysb.2004.11.048} {\bibfield
  {journal} {\bibinfo  {journal} {Nuclear Physics B}\ }\textbf {\bibinfo
  {volume} {706}},\ \bibinfo {pages} {487} (\bibinfo {year} {2005})},\ \bibinfo
  {note} {arXiv: hep-ph/0406088}\BibitemShut {NoStop}%
\bibitem [{\citenamefont {Randall}\ and\ \citenamefont
  {Sundrum}(1999)}]{randall_out_1999}%
  \BibitemOpen
  \bibfield  {author} {\bibinfo {author} {\bibfnamefont {L.}~\bibnamefont
  {Randall}}\ and\ \bibinfo {author} {\bibfnamefont {R.}~\bibnamefont
  {Sundrum}},\ }\href {\doibase 10.1016/S0550-3213(99)00359-4} {\bibfield
  {journal} {\bibinfo  {journal} {Nuclear Physics B}\ }\textbf {\bibinfo
  {volume} {557}} (\bibinfo {year} {1999}),\
  10.1016/S0550-3213(99)00359-4}\BibitemShut {NoStop}%
\bibitem [{\citenamefont {Giudice}\ \emph {et~al.}(1998)\citenamefont
  {Giudice}, \citenamefont {Luty}, \citenamefont {Murayama},\ and\
  \citenamefont {Rattazzi}}]{Giudice:1998xp}%
  \BibitemOpen
  \bibfield  {author} {\bibinfo {author} {\bibfnamefont {G.~F.}\ \bibnamefont
  {Giudice}}, \bibinfo {author} {\bibfnamefont {M.~A.}\ \bibnamefont {Luty}},
  \bibinfo {author} {\bibfnamefont {H.}~\bibnamefont {Murayama}}, \ and\
  \bibinfo {author} {\bibfnamefont {R.}~\bibnamefont {Rattazzi}},\ }\href
  {\doibase 10.1088/1126-6708/1998/12/027} {\bibfield  {journal} {\bibinfo
  {journal} {JHEP}\ }\textbf {\bibinfo {volume} {12}},\ \bibinfo {pages} {027}
  (\bibinfo {year} {1998})},\ \Eprint {http://arxiv.org/abs/hep-ph/9810442}
  {arXiv:hep-ph/9810442} \BibitemShut {NoStop}%
\bibitem [{\citenamefont {Aaboud}\ \emph {et~al.}(2018)\citenamefont {Aaboud}
  \emph {et~al.}}]{Aaboud:2017leg}%
  \BibitemOpen
  \bibfield  {author} {\bibinfo {author} {\bibfnamefont {M.}~\bibnamefont
  {Aaboud}} \emph {et~al.} (\bibinfo {collaboration} {ATLAS}),\ }\href
  {\doibase 10.1103/PhysRevD.97.052010} {\bibfield  {journal} {\bibinfo
  {journal} {Phys. Rev. D}\ }\textbf {\bibinfo {volume} {97}},\ \bibinfo
  {pages} {052010} (\bibinfo {year} {2018})},\ \Eprint
  {http://arxiv.org/abs/1712.08119} {arXiv:1712.08119 [hep-ex]} \BibitemShut
  {NoStop}%
\bibitem [{\citenamefont {Arkani-Hamed}\ \emph {et~al.}(2005)\citenamefont
  {Arkani-Hamed}, \citenamefont {Dimopoulos}, \citenamefont {Giudice},\ and\
  \citenamefont {Romanino}}]{arkani-hamed_aspects_2005}%
  \BibitemOpen
  \bibfield  {author} {\bibinfo {author} {\bibfnamefont {N.}~\bibnamefont
  {Arkani-Hamed}}, \bibinfo {author} {\bibfnamefont {S.}~\bibnamefont
  {Dimopoulos}}, \bibinfo {author} {\bibfnamefont {G.~F.}\ \bibnamefont
  {Giudice}}, \ and\ \bibinfo {author} {\bibfnamefont {A.}~\bibnamefont
  {Romanino}},\ }\href {\doibase 10.1016/j.nuclphysb.2004.12.026} {\bibfield
  {journal} {\bibinfo  {journal} {Nuclear Physics B}\ }\textbf {\bibinfo
  {volume} {709}} (\bibinfo {year} {2005}),\
  10.1016/j.nuclphysb.2004.12.026}\BibitemShut {NoStop}%
\bibitem [{\citenamefont {Baer}\ \emph
  {et~al.}(2012{\natexlab{a}})\citenamefont {Baer}, \citenamefont {Barger},
  \citenamefont {Huang},\ and\ \citenamefont {Tata}}]{Baer:2012uy}%
  \BibitemOpen
  \bibfield  {author} {\bibinfo {author} {\bibfnamefont {H.}~\bibnamefont
  {Baer}}, \bibinfo {author} {\bibfnamefont {V.}~\bibnamefont {Barger}},
  \bibinfo {author} {\bibfnamefont {P.}~\bibnamefont {Huang}}, \ and\ \bibinfo
  {author} {\bibfnamefont {X.}~\bibnamefont {Tata}},\ }\href {\doibase
  10.1007/JHEP05(2012)109} {\bibfield  {journal} {\bibinfo  {journal} {JHEP}\
  }\textbf {\bibinfo {volume} {05}},\ \bibinfo {pages} {109} (\bibinfo {year}
  {2012}{\natexlab{a}})},\ \Eprint {http://arxiv.org/abs/1203.5539}
  {arXiv:1203.5539 [hep-ph]} \BibitemShut {NoStop}%
\bibitem [{\citenamefont {Baer}\ \emph
  {et~al.}(2012{\natexlab{b}})\citenamefont {Baer}, \citenamefont {Barger},
  \citenamefont {Huang}, \citenamefont {Mustafayev},\ and\ \citenamefont
  {Tata}}]{Baer:2012up}%
  \BibitemOpen
  \bibfield  {author} {\bibinfo {author} {\bibfnamefont {H.}~\bibnamefont
  {Baer}}, \bibinfo {author} {\bibfnamefont {V.}~\bibnamefont {Barger}},
  \bibinfo {author} {\bibfnamefont {P.}~\bibnamefont {Huang}}, \bibinfo
  {author} {\bibfnamefont {A.}~\bibnamefont {Mustafayev}}, \ and\ \bibinfo
  {author} {\bibfnamefont {X.}~\bibnamefont {Tata}},\ }\href {\doibase
  10.1103/PhysRevLett.109.161802} {\bibfield  {journal} {\bibinfo  {journal}
  {Phys. Rev. Lett.}\ }\textbf {\bibinfo {volume} {109}},\ \bibinfo {pages}
  {161802} (\bibinfo {year} {2012}{\natexlab{b}})},\ \Eprint
  {http://arxiv.org/abs/1207.3343} {arXiv:1207.3343 [hep-ph]} \BibitemShut
  {NoStop}%
\bibitem [{\citenamefont {Arkani-Hamed}\ and\ \citenamefont
  {Dimopoulos}(2005)}]{ArkaniHame_supersymmetric_2005}%
  \BibitemOpen
  \bibfield  {author} {\bibinfo {author} {\bibfnamefont {N.}~\bibnamefont
  {Arkani-Hamed}}\ and\ \bibinfo {author} {\bibfnamefont {S.}~\bibnamefont
  {Dimopoulos}},\ }\href {\doibase 10.1088/1126-6708/2005/06/073} {\bibfield
  {journal} {\bibinfo  {journal} {JHEP}\ }\textbf {\bibinfo {volume} {06}},\
  \bibinfo {pages} {073} (\bibinfo {year} {2005})},\ \Eprint
  {http://arxiv.org/abs/hep-th/0405159} {arXiv:hep-th/0405159} \BibitemShut
  {NoStop}%
\bibitem [{\citenamefont {Arbey}\ \emph {et~al.}(2012)\citenamefont {Arbey},
  \citenamefont {Battaglia}, \citenamefont {Djouadi}, \citenamefont
  {Mahmoudi},\ and\ \citenamefont {Quevillon}}]{Arbey_Implication_2012}%
  \BibitemOpen
  \bibfield  {author} {\bibinfo {author} {\bibfnamefont {A.}~\bibnamefont
  {Arbey}}, \bibinfo {author} {\bibfnamefont {M.}~\bibnamefont {Battaglia}},
  \bibinfo {author} {\bibfnamefont {A.}~\bibnamefont {Djouadi}}, \bibinfo
  {author} {\bibfnamefont {F.}~\bibnamefont {Mahmoudi}}, \ and\ \bibinfo
  {author} {\bibfnamefont {J.}~\bibnamefont {Quevillon}},\ }\href {\doibase
  10.1016/j.physletb.2012.01.053} {\bibfield  {journal} {\bibinfo  {journal}
  {Phys. Lett. B}\ }\textbf {\bibinfo {volume} {708}},\ \bibinfo {pages} {162}
  (\bibinfo {year} {2012})},\ \Eprint {http://arxiv.org/abs/1112.3028}
  {arXiv:1112.3028 [hep-ph]} \BibitemShut {NoStop}%
\bibitem [{\citenamefont {Wells}(2005)}]{wells_pev-scale_2005}%
  \BibitemOpen
  \bibfield  {author} {\bibinfo {author} {\bibfnamefont {J.~D.}\ \bibnamefont
  {Wells}},\ }\href {\doibase 10.1103/PhysRevD.71.015013} {\bibfield  {journal}
  {\bibinfo  {journal} {Physical Review D}\ }\textbf {\bibinfo {volume} {71}},\
  \bibinfo {pages} {015013} (\bibinfo {year} {2005})},\ \bibinfo {note} {arXiv:
  hep-ph/0411041}\BibitemShut {NoStop}%
\bibitem [{\citenamefont {Giudice}\ and\ \citenamefont
  {Romanino}(2006)}]{giudice_electric_2006}%
  \BibitemOpen
  \bibfield  {author} {\bibinfo {author} {\bibfnamefont {G.~F.}\ \bibnamefont
  {Giudice}}\ and\ \bibinfo {author} {\bibfnamefont {A.}~\bibnamefont
  {Romanino}},\ }\href {\doibase 10.1016/j.physletb.2006.01.027} {\bibfield
  {journal} {\bibinfo  {journal} {Physics Letters B}\ }\textbf {\bibinfo
  {volume} {634}},\ \bibinfo {pages} {307} (\bibinfo {year} {2006})},\ \bibinfo
  {note} {arXiv: hep-ph/0510197}\BibitemShut {NoStop}%
\bibitem [{\citenamefont {Alloul}\ \emph {et~al.}(2014)\citenamefont {Alloul},
  \citenamefont {Christensen}, \citenamefont {Degrande}, \citenamefont {Duhr},\
  and\ \citenamefont {Fuks}}]{feynrules}%
  \BibitemOpen
  \bibfield  {author} {\bibinfo {author} {\bibfnamefont {A.}~\bibnamefont
  {Alloul}}, \bibinfo {author} {\bibfnamefont {N.~D.}\ \bibnamefont
  {Christensen}}, \bibinfo {author} {\bibfnamefont {C.}~\bibnamefont
  {Degrande}}, \bibinfo {author} {\bibfnamefont {C.}~\bibnamefont {Duhr}}, \
  and\ \bibinfo {author} {\bibfnamefont {B.}~\bibnamefont {Fuks}},\ }\href
  {\doibase 10.1016/j.cpc.2014.04.012} {\bibfield  {journal} {\bibinfo
  {journal} {Comput. Phys. Commun.}\ }\textbf {\bibinfo {volume} {185}},\
  \bibinfo {pages} {2250} (\bibinfo {year} {2014})},\ \Eprint
  {http://arxiv.org/abs/1310.1921} {arXiv:1310.1921 [hep-ph]} \BibitemShut
  {NoStop}%
\bibitem [{\citenamefont {B\'elanger}\ \emph {et~al.}(2018)\citenamefont
  {B\'elanger}, \citenamefont {Boudjema}, \citenamefont {Goudelis},
  \citenamefont {Pukhov},\ and\ \citenamefont {Zaldivar}}]{micromegas}%
  \BibitemOpen
  \bibfield  {author} {\bibinfo {author} {\bibfnamefont {G.}~\bibnamefont
  {B\'elanger}}, \bibinfo {author} {\bibfnamefont {F.}~\bibnamefont
  {Boudjema}}, \bibinfo {author} {\bibfnamefont {A.}~\bibnamefont {Goudelis}},
  \bibinfo {author} {\bibfnamefont {A.}~\bibnamefont {Pukhov}}, \ and\ \bibinfo
  {author} {\bibfnamefont {B.}~\bibnamefont {Zaldivar}},\ }\href {\doibase
  10.1016/j.cpc.2018.04.027} {\bibfield  {journal} {\bibinfo  {journal}
  {Comput. Phys. Commun.}\ }\textbf {\bibinfo {volume} {231}},\ \bibinfo
  {pages} {173} (\bibinfo {year} {2018})},\ \Eprint
  {http://arxiv.org/abs/1801.03509} {arXiv:1801.03509 [hep-ph]} \BibitemShut
  {NoStop}%
\bibitem [{\citenamefont {Aprile}\ \emph {et~al.}(2018)\citenamefont {Aprile}
  \emph {et~al.}}]{aprile_dark_2018}%
  \BibitemOpen
  \bibfield  {author} {\bibinfo {author} {\bibfnamefont {E.}~\bibnamefont
  {Aprile}} \emph {et~al.} (\bibinfo {collaboration} {XENON}),\ }\href
  {\doibase 10.1103/PhysRevLett.121.111302} {\bibfield  {journal} {\bibinfo
  {journal} {Phys. Rev. Lett.}\ }\textbf {\bibinfo {volume} {121}},\ \bibinfo
  {pages} {111302} (\bibinfo {year} {2018})},\ \Eprint
  {http://arxiv.org/abs/1805.12562} {arXiv:1805.12562 [astro-ph.CO]}
  \BibitemShut {NoStop}%
\bibitem [{\citenamefont {Cautun}\ \emph {et~al.}(2020)\citenamefont {Cautun},
  \citenamefont {Benítez-Llambay}, \citenamefont {Deason}, \citenamefont
  {Frenk}, \citenamefont {Fattahi}, \citenamefont {Gómez}, \citenamefont
  {Grand}, \citenamefont {Oman}, \citenamefont {Navarro},\ and\ \citenamefont
  {Simpson}}]{Cautun_milky_2020}%
  \BibitemOpen
  \bibfield  {author} {\bibinfo {author} {\bibfnamefont {M.}~\bibnamefont
  {Cautun}}, \bibinfo {author} {\bibfnamefont {A.}~\bibnamefont
  {Benítez-Llambay}}, \bibinfo {author} {\bibfnamefont {A.~J.}\ \bibnamefont
  {Deason}}, \bibinfo {author} {\bibfnamefont {C.~S.}\ \bibnamefont {Frenk}},
  \bibinfo {author} {\bibfnamefont {A.}~\bibnamefont {Fattahi}}, \bibinfo
  {author} {\bibfnamefont {F.~A.}\ \bibnamefont {Gómez}}, \bibinfo {author}
  {\bibfnamefont {R.~J.~J.}\ \bibnamefont {Grand}}, \bibinfo {author}
  {\bibfnamefont {K.~A.}\ \bibnamefont {Oman}}, \bibinfo {author}
  {\bibfnamefont {J.~F.}\ \bibnamefont {Navarro}}, \ and\ \bibinfo {author}
  {\bibfnamefont {C.~M.}\ \bibnamefont {Simpson}},\ }\href {\doibase
  10.1093/mnras/staa1017} {\bibfield  {journal} {\bibinfo  {journal} {Monthly
  Notices of the Royal Astronomical Society}\ }\textbf {\bibinfo {volume}
  {494}},\ \bibinfo {pages} {4291–4313} (\bibinfo {year} {2020})}\BibitemShut
  {NoStop}%
\bibitem [{\citenamefont {Nitschai}\ \emph {et~al.}(2020)\citenamefont
  {Nitschai}, \citenamefont {Cappellari},\ and\ \citenamefont
  {Neumayer}}]{Nitschai_First-Gaia_2020}%
  \BibitemOpen
  \bibfield  {author} {\bibinfo {author} {\bibfnamefont {M.~S.}\ \bibnamefont
  {Nitschai}}, \bibinfo {author} {\bibfnamefont {M.}~\bibnamefont
  {Cappellari}}, \ and\ \bibinfo {author} {\bibfnamefont {N.}~\bibnamefont
  {Neumayer}},\ }\href {\doibase 10.1093/mnras/staa1128} {\bibfield  {journal}
  {\bibinfo  {journal} {Monthly Notices of the Royal Astronomical Society}\
  }\textbf {\bibinfo {volume} {494}},\ \bibinfo {pages} {6001–6011} (\bibinfo
  {year} {2020})}\BibitemShut {NoStop}%
\bibitem [{\citenamefont {Jungman}\ \emph {et~al.}(1996)\citenamefont
  {Jungman}, \citenamefont {Kamionkowski},\ and\ \citenamefont
  {Griest}}]{jungman_supersymmetric_1996}%
  \BibitemOpen
  \bibfield  {author} {\bibinfo {author} {\bibfnamefont {G.}~\bibnamefont
  {Jungman}}, \bibinfo {author} {\bibfnamefont {M.}~\bibnamefont
  {Kamionkowski}}, \ and\ \bibinfo {author} {\bibfnamefont {K.}~\bibnamefont
  {Griest}},\ }\href {\doibase 10.1016/0370-1573(95)00058-5} {\bibfield
  {journal} {\bibinfo  {journal} {Physics Reports}\ }\textbf {\bibinfo {volume}
  {267}},\ \bibinfo {pages} {195} (\bibinfo {year} {1996})}\BibitemShut
  {NoStop}%
\bibitem [{may(2016)}]{mayet_review_2016}%
  \BibitemOpen
  \href {\doibase 10.1016/j.physrep.2016.02.007} {\ \textbf {\bibinfo {volume}
  {627}},\ \bibinfo {pages} {1} (\bibinfo {year} {2016})},\ \bibinfo {note}
  {arXiv: 1602.03781}\BibitemShut {NoStop}%
\bibitem [{\citenamefont {Belanger}\ \emph {et~al.}(2009)\citenamefont
  {Belanger}, \citenamefont {Nezri},\ and\ \citenamefont
  {Pukhov}}]{Belanger_Discriminating_2009}%
  \BibitemOpen
  \bibfield  {author} {\bibinfo {author} {\bibfnamefont {G.}~\bibnamefont
  {Belanger}}, \bibinfo {author} {\bibfnamefont {E.}~\bibnamefont {Nezri}}, \
  and\ \bibinfo {author} {\bibfnamefont {A.}~\bibnamefont {Pukhov}},\ }\href
  {\doibase 10.1103/PhysRevD.79.015008} {\bibfield  {journal} {\bibinfo
  {journal} {Phys. Rev. D}\ }\textbf {\bibinfo {volume} {79}},\ \bibinfo
  {pages} {015008} (\bibinfo {year} {2009})},\ \Eprint
  {http://arxiv.org/abs/0810.1362} {arXiv:0810.1362 [hep-ph]} \BibitemShut
  {NoStop}%
\bibitem [{\citenamefont {Cohen}\ \emph {et~al.}(2010)\citenamefont {Cohen},
  \citenamefont {Phalen},\ and\ \citenamefont
  {Pierce}}]{Cohen_Correlation_2010}%
  \BibitemOpen
  \bibfield  {author} {\bibinfo {author} {\bibfnamefont {T.}~\bibnamefont
  {Cohen}}, \bibinfo {author} {\bibfnamefont {D.~J.}\ \bibnamefont {Phalen}}, \
  and\ \bibinfo {author} {\bibfnamefont {A.}~\bibnamefont {Pierce}},\ }\href
  {\doibase 10.1103/PhysRevD.81.116001} {\bibfield  {journal} {\bibinfo
  {journal} {Phys. Rev. D}\ }\textbf {\bibinfo {volume} {81}},\ \bibinfo
  {pages} {116001} (\bibinfo {year} {2010})},\ \Eprint
  {http://arxiv.org/abs/1001.3408} {arXiv:1001.3408 [hep-ph]} \BibitemShut
  {NoStop}%
\bibitem [{\citenamefont {Cheung}\ \emph {et~al.}(2013)\citenamefont {Cheung},
  \citenamefont {Hall}, \citenamefont {Pinner},\ and\ \citenamefont
  {Ruderman}}]{cheung_prospects_2013}%
  \BibitemOpen
  \bibfield  {author} {\bibinfo {author} {\bibfnamefont {C.}~\bibnamefont
  {Cheung}}, \bibinfo {author} {\bibfnamefont {L.~J.}\ \bibnamefont {Hall}},
  \bibinfo {author} {\bibfnamefont {D.}~\bibnamefont {Pinner}}, \ and\ \bibinfo
  {author} {\bibfnamefont {J.~T.}\ \bibnamefont {Ruderman}},\ }\href {\doibase
  10.1007/JHEP05(2013)100} {\bibfield  {journal} {\bibinfo  {journal} {Journal
  of High Energy Physics}\ }\textbf {\bibinfo {volume} {2013}},\ \bibinfo
  {pages} {100} (\bibinfo {year} {2013})},\ \bibinfo {note} {arXiv:
  1211.4873}\BibitemShut {NoStop}%
\bibitem [{\citenamefont {Hisano}\ \emph {et~al.}(2011)\citenamefont {Hisano},
  \citenamefont {Ishiwata}, \citenamefont {Nagata},\ and\ \citenamefont
  {Takesako}}]{hisano_direct_2011}%
  \BibitemOpen
  \bibfield  {author} {\bibinfo {author} {\bibfnamefont {J.}~\bibnamefont
  {Hisano}}, \bibinfo {author} {\bibfnamefont {K.}~\bibnamefont {Ishiwata}},
  \bibinfo {author} {\bibfnamefont {N.}~\bibnamefont {Nagata}}, \ and\ \bibinfo
  {author} {\bibfnamefont {T.}~\bibnamefont {Takesako}},\ }\href {\doibase
  10.1007/JHEP07(2011)005} {\ \textbf {\bibinfo {volume} {2011}},\ \bibinfo
  {pages} {5} (\bibinfo {year} {2011})},\ \bibinfo {note} {arXiv:
  1104.0228}\BibitemShut {NoStop}%
\bibitem [{\citenamefont {Hisano}\ \emph {et~al.}(2013)\citenamefont {Hisano},
  \citenamefont {Ishiwata},\ and\ \citenamefont {Nagata}}]{hisano_direct_2013}%
  \BibitemOpen
  \bibfield  {author} {\bibinfo {author} {\bibfnamefont {J.}~\bibnamefont
  {Hisano}}, \bibinfo {author} {\bibfnamefont {K.}~\bibnamefont {Ishiwata}}, \
  and\ \bibinfo {author} {\bibfnamefont {N.}~\bibnamefont {Nagata}},\ }\href
  {\doibase 10.1103/PhysRevD.87.035020} {\bibfield  {journal} {\bibinfo
  {journal} {Physical Review D}\ }\textbf {\bibinfo {volume} {87}},\ \bibinfo
  {pages} {035020} (\bibinfo {year} {2013})},\ \bibinfo {note} {arXiv:
  1210.5985}\BibitemShut {NoStop}%
\bibitem [{\citenamefont {Akerib}\ \emph {et~al.}(2020)\citenamefont {Akerib}
  \emph {et~al.}}]{LZ_projected_2020}%
  \BibitemOpen
  \bibfield  {author} {\bibinfo {author} {\bibfnamefont {D.~S.}\ \bibnamefont
  {Akerib}} \emph {et~al.} (\bibinfo {collaboration} {LUX-ZEPLIN}),\ }\href
  {\doibase 10.1103/PhysRevD.101.052002} {\bibfield  {journal} {\bibinfo
  {journal} {Physical Review D}\ }\textbf {\bibinfo {volume} {101}},\ \bibinfo
  {pages} {052002} (\bibinfo {year} {2020})},\ \bibinfo {note} {arXiv:
  1802.06039}\BibitemShut {NoStop}%
\bibitem [{\citenamefont {Aprile}\ \emph {et~al.}(2017)\citenamefont {Aprile}
  \emph {et~al.}}]{xenon_1T_original}%
  \BibitemOpen
  \bibfield  {author} {\bibinfo {author} {\bibfnamefont {E.}~\bibnamefont
  {Aprile}} \emph {et~al.} (\bibinfo {collaboration} {XENON}),\ }\href
  {\doibase 10.1103/PhysRevLett.119.181301} {\bibfield  {journal} {\bibinfo
  {journal} {Phys. Rev. Lett.}\ }\textbf {\bibinfo {volume} {119}},\ \bibinfo
  {pages} {181301} (\bibinfo {year} {2017})},\ \Eprint
  {http://arxiv.org/abs/1705.06655} {arXiv:1705.06655 [astro-ph.CO]}
  \BibitemShut {NoStop}%
\bibitem [{\citenamefont {Slatyer}(2017)}]{slatyer_tasi_2017}%
  \BibitemOpen
  \bibfield  {author} {\bibinfo {author} {\bibfnamefont {T.~R.}\ \bibnamefont
  {Slatyer}},\ }\href {http://arxiv.org/abs/1710.05137} {\bibfield  {journal}
  {\bibinfo  {journal} {arXiv:1710.05137 [astro-ph, physics:hep-ph]}\ }
  (\bibinfo {year} {2017})},\ \bibinfo {note} {arXiv: 1710.05137}\BibitemShut
  {NoStop}%
\bibitem [{\citenamefont {Actis}\ \emph {et~al.}(2011)\citenamefont {Actis}
  \emph {et~al.}}]{CTA_Ground_2011}%
  \BibitemOpen
  \bibfield  {author} {\bibinfo {author} {\bibfnamefont {M.}~\bibnamefont
  {Actis}} \emph {et~al.} (\bibinfo {collaboration} {CTA Consortium}),\ }\href
  {\doibase 10.1007/s10686-011-9247-0} {\bibfield  {journal} {\bibinfo
  {journal} {Experimental Astronomy}\ }\textbf {\bibinfo {volume} {32}},\
  \bibinfo {pages} {193} (\bibinfo {year} {2011})},\ \Eprint
  {http://arxiv.org/abs/1008.3703} {arXiv:1008.3703 [astro-ph.IM]} \BibitemShut
  {NoStop}%
\bibitem [{\citenamefont {Hryczuk}\ \emph {et~al.}(2019)\citenamefont
  {Hryczuk}, \citenamefont {Jodłowski}, \citenamefont {Moulin}, \citenamefont
  {Rinchiuso}, \citenamefont {Roszkowski}, \citenamefont {Sessolo},\ and\
  \citenamefont {Trojanowski}}]{hryczuk_testing_2019}%
  \BibitemOpen
  \bibfield  {author} {\bibinfo {author} {\bibfnamefont {A.}~\bibnamefont
  {Hryczuk}}, \bibinfo {author} {\bibfnamefont {K.}~\bibnamefont {Jodłowski}},
  \bibinfo {author} {\bibfnamefont {E.}~\bibnamefont {Moulin}}, \bibinfo
  {author} {\bibfnamefont {L.}~\bibnamefont {Rinchiuso}}, \bibinfo {author}
  {\bibfnamefont {L.}~\bibnamefont {Roszkowski}}, \bibinfo {author}
  {\bibfnamefont {E.~M.}\ \bibnamefont {Sessolo}}, \ and\ \bibinfo {author}
  {\bibfnamefont {S.}~\bibnamefont {Trojanowski}},\ }\href {\doibase
  10.1007/JHEP10(2019)043} {\bibfield  {journal} {\bibinfo  {journal} {Journal
  of High Energy Physics}\ }\textbf {\bibinfo {volume} {2019}},\ \bibinfo
  {pages} {43} (\bibinfo {year} {2019})},\ \bibinfo {note} {arXiv:
  1905.00315}\BibitemShut {NoStop}%
\bibitem [{\citenamefont {de~Blok}(2010)}]{de_blok_core-cusp_2010}%
  \BibitemOpen
  \bibfield  {author} {\bibinfo {author} {\bibfnamefont {W.~J.~G.}\
  \bibnamefont {de~Blok}},\ }\href {\doibase 10.1155/2010/789293} {\bibfield
  {journal} {\bibinfo  {journal} {Advances in Astronomy}\ }\textbf {\bibinfo
  {volume} {2010}},\ \bibinfo {pages} {1} (\bibinfo {year} {2010})},\ \bibinfo
  {note} {arXiv: 0910.3538}\BibitemShut {NoStop}%
\bibitem [{\citenamefont {Einasto}(1965)}]{Einasto:1965czb}%
  \BibitemOpen
  \bibfield  {author} {\bibinfo {author} {\bibfnamefont {J.}~\bibnamefont
  {Einasto}},\ }\href@noop {} {\bibfield  {journal} {\bibinfo  {journal} {Trudy
  Astrofizicheskogo Instituta Alma-Ata}\ }\textbf {\bibinfo {volume} {5}},\
  \bibinfo {pages} {87} (\bibinfo {year} {1965})}\BibitemShut {NoStop}%
\bibitem [{\citenamefont {Navarro}\ \emph {et~al.}(1996)\citenamefont
  {Navarro}, \citenamefont {Frenk},\ and\ \citenamefont
  {White}}]{Navarro_Structure_1996}%
  \BibitemOpen
  \bibfield  {author} {\bibinfo {author} {\bibfnamefont {J.~F.}\ \bibnamefont
  {Navarro}}, \bibinfo {author} {\bibfnamefont {C.~S.}\ \bibnamefont {Frenk}},
  \ and\ \bibinfo {author} {\bibfnamefont {S.~D.~M.}\ \bibnamefont {White}},\
  }\href {\doibase 10.1086/177173} {\bibfield  {journal} {\bibinfo  {journal}
  {Astrophys. J.}\ }\textbf {\bibinfo {volume} {462}},\ \bibinfo {pages} {563}
  (\bibinfo {year} {1996})},\ \Eprint {http://arxiv.org/abs/astro-ph/9508025}
  {arXiv:astro-ph/9508025} \BibitemShut {NoStop}%
\bibitem [{\citenamefont {Pieri}\ \emph {et~al.}(2011)\citenamefont {Pieri},
  \citenamefont {Lavalle}, \citenamefont {Bertone},\ and\ \citenamefont
  {Branchini}}]{Pieri_Implications_2011}%
  \BibitemOpen
  \bibfield  {author} {\bibinfo {author} {\bibfnamefont {L.}~\bibnamefont
  {Pieri}}, \bibinfo {author} {\bibfnamefont {J.}~\bibnamefont {Lavalle}},
  \bibinfo {author} {\bibfnamefont {G.}~\bibnamefont {Bertone}}, \ and\
  \bibinfo {author} {\bibfnamefont {E.}~\bibnamefont {Branchini}},\ }\href
  {\doibase 10.1103/PhysRevD.83.023518} {\bibfield  {journal} {\bibinfo
  {journal} {Phys. Rev. D}\ }\textbf {\bibinfo {volume} {83}},\ \bibinfo
  {pages} {023518} (\bibinfo {year} {2011})},\ \Eprint
  {http://arxiv.org/abs/0908.0195} {arXiv:0908.0195 [astro-ph.HE]} \BibitemShut
  {NoStop}%
\bibitem [{\citenamefont {Abramowski}\ \emph {et~al.}(2011)\citenamefont
  {Abramowski} \emph {et~al.}}]{Abramowski_Search_2011}%
  \BibitemOpen
  \bibfield  {author} {\bibinfo {author} {\bibfnamefont {A.}~\bibnamefont
  {Abramowski}} \emph {et~al.} (\bibinfo {collaboration} {H.E.S.S.}),\ }\href
  {\doibase 10.1103/PhysRevLett.106.161301} {\bibfield  {journal} {\bibinfo
  {journal} {Phys. Rev. Lett.}\ }\textbf {\bibinfo {volume} {106}},\ \bibinfo
  {pages} {161301} (\bibinfo {year} {2011})},\ \Eprint
  {http://arxiv.org/abs/1103.3266} {arXiv:1103.3266 [astro-ph.HE]} \BibitemShut
  {NoStop}%
\bibitem [{\citenamefont {Rinchiuso}\ \emph {et~al.}(2021)\citenamefont
  {Rinchiuso}, \citenamefont {Macias}, \citenamefont {Moulin}, \citenamefont
  {Rodd},\ and\ \citenamefont {Slatyer}}]{slatyer_prospects_2021}%
  \BibitemOpen
  \bibfield  {author} {\bibinfo {author} {\bibfnamefont {L.}~\bibnamefont
  {Rinchiuso}}, \bibinfo {author} {\bibfnamefont {O.}~\bibnamefont {Macias}},
  \bibinfo {author} {\bibfnamefont {E.}~\bibnamefont {Moulin}}, \bibinfo
  {author} {\bibfnamefont {N.~L.}\ \bibnamefont {Rodd}}, \ and\ \bibinfo
  {author} {\bibfnamefont {T.~R.}\ \bibnamefont {Slatyer}},\ }\href {\doibase
  10.1103/PhysRevD.103.023011} {\bibfield  {journal} {\bibinfo  {journal}
  {Physical Review D}\ }\textbf {\bibinfo {volume} {103}},\ \bibinfo {pages}
  {023011} (\bibinfo {year} {2021})},\ \bibinfo {note} {arXiv:
  2008.00692}\BibitemShut {NoStop}%
\bibitem [{\citenamefont {Hisano}\ \emph {et~al.}(2004)\citenamefont {Hisano},
  \citenamefont {Matsumoto},\ and\ \citenamefont {Nojiri}}]{Hisano:2003ec}%
  \BibitemOpen
  \bibfield  {author} {\bibinfo {author} {\bibfnamefont {J.}~\bibnamefont
  {Hisano}}, \bibinfo {author} {\bibfnamefont {S.}~\bibnamefont {Matsumoto}}, \
  and\ \bibinfo {author} {\bibfnamefont {M.~M.}\ \bibnamefont {Nojiri}},\
  }\href {\doibase 10.1103/PhysRevLett.92.031303} {\bibfield  {journal}
  {\bibinfo  {journal} {Phys. Rev. Lett.}\ }\textbf {\bibinfo {volume} {92}},\
  \bibinfo {pages} {031303} (\bibinfo {year} {2004})},\ \Eprint
  {http://arxiv.org/abs/hep-ph/0307216} {arXiv:hep-ph/0307216} \BibitemShut
  {NoStop}%
\bibitem [{\citenamefont {Hisano}\ \emph {et~al.}(2005)\citenamefont {Hisano},
  \citenamefont {Matsumoto}, \citenamefont {Nojiri},\ and\ \citenamefont
  {Saito}}]{Hisano:2004ds}%
  \BibitemOpen
  \bibfield  {author} {\bibinfo {author} {\bibfnamefont {J.}~\bibnamefont
  {Hisano}}, \bibinfo {author} {\bibfnamefont {S.}~\bibnamefont {Matsumoto}},
  \bibinfo {author} {\bibfnamefont {M.~M.}\ \bibnamefont {Nojiri}}, \ and\
  \bibinfo {author} {\bibfnamefont {O.}~\bibnamefont {Saito}},\ }\href
  {\doibase 10.1103/PhysRevD.71.063528} {\bibfield  {journal} {\bibinfo
  {journal} {Phys. Rev. D}\ }\textbf {\bibinfo {volume} {71}},\ \bibinfo
  {pages} {063528} (\bibinfo {year} {2005})},\ \Eprint
  {http://arxiv.org/abs/hep-ph/0412403} {arXiv:hep-ph/0412403} \BibitemShut
  {NoStop}%
\bibitem [{\citenamefont {Krall}\ and\ \citenamefont
  {Reece}(2018)}]{krall_last_2018}%
  \BibitemOpen
  \bibfield  {author} {\bibinfo {author} {\bibfnamefont {R.}~\bibnamefont
  {Krall}}\ and\ \bibinfo {author} {\bibfnamefont {M.}~\bibnamefont {Reece}},\
  }\href {\doibase 10.1088/1674-1137/42/4/043105} {\bibfield  {journal}
  {\bibinfo  {journal} {Chinese Physics C}\ }\textbf {\bibinfo {volume} {42}}
  (\bibinfo {year} {2018}),\ 10.1088/1674-1137/42/4/043105},\ \bibinfo {note}
  {arXiv: 1705.04843}\BibitemShut {NoStop}%
\bibitem [{\citenamefont {Galli}\ \emph {et~al.}(2009)\citenamefont {Galli},
  \citenamefont {Iocco}, \citenamefont {Bertone},\ and\ \citenamefont
  {Melchiorri}}]{galli_cmb_2009}%
  \BibitemOpen
  \bibfield  {author} {\bibinfo {author} {\bibfnamefont {S.}~\bibnamefont
  {Galli}}, \bibinfo {author} {\bibfnamefont {F.}~\bibnamefont {Iocco}},
  \bibinfo {author} {\bibfnamefont {G.}~\bibnamefont {Bertone}}, \ and\
  \bibinfo {author} {\bibfnamefont {A.}~\bibnamefont {Melchiorri}},\ }\href
  {\doibase 10.1103/PhysRevD.80.023505} {\bibfield  {journal} {\bibinfo
  {journal} {Physical Review D}\ }\textbf {\bibinfo {volume} {80}},\ \bibinfo
  {pages} {023505} (\bibinfo {year} {2009})},\ \bibinfo {note} {arXiv:
  0905.0003}\BibitemShut {NoStop}%
\bibitem [{\citenamefont {Madhavacheril}\ \emph {et~al.}(2014)\citenamefont
  {Madhavacheril}, \citenamefont {Sehgal},\ and\ \citenamefont
  {Slatyer}}]{madhavacheril_current_2014}%
  \BibitemOpen
  \bibfield  {author} {\bibinfo {author} {\bibfnamefont {M.~S.}\ \bibnamefont
  {Madhavacheril}}, \bibinfo {author} {\bibfnamefont {N.}~\bibnamefont
  {Sehgal}}, \ and\ \bibinfo {author} {\bibfnamefont {T.~R.}\ \bibnamefont
  {Slatyer}},\ }\href {\doibase 10.1103/PhysRevD.89.103508} {\bibfield
  {journal} {\bibinfo  {journal} {Physical Review D}\ }\textbf {\bibinfo
  {volume} {89}},\ \bibinfo {pages} {103508} (\bibinfo {year} {2014})},\
  \bibinfo {note} {arXiv: 1310.3815}\BibitemShut {NoStop}%
\bibitem [{\citenamefont {Andreev}\ \emph {et~al.}(2018)\citenamefont {Andreev}
  \emph {et~al.}}]{ACME_improved_2018}%
  \BibitemOpen
  \bibfield  {author} {\bibinfo {author} {\bibfnamefont {V.}~\bibnamefont
  {Andreev}} \emph {et~al.} (\bibinfo {collaboration} {ACME}),\ }\href
  {\doibase 10.1038/s41586-018-0599-8} {\bibfield  {journal} {\bibinfo
  {journal} {Nature}\ }\textbf {\bibinfo {volume} {562}},\ \bibinfo {pages}
  {355} (\bibinfo {year} {2018})}\BibitemShut {NoStop}%
\bibitem [{\citenamefont {Pospelov}\ and\ \citenamefont
  {Ritz}(2014)}]{Pospelov_CKM_2014}%
  \BibitemOpen
  \bibfield  {author} {\bibinfo {author} {\bibfnamefont {M.}~\bibnamefont
  {Pospelov}}\ and\ \bibinfo {author} {\bibfnamefont {A.}~\bibnamefont
  {Ritz}},\ }\href {\doibase 10.1103/PhysRevD.89.056006} {\bibfield  {journal}
  {\bibinfo  {journal} {Phys. Rev. D}\ }\textbf {\bibinfo {volume} {89}},\
  \bibinfo {pages} {056006} (\bibinfo {year} {2014})},\ \Eprint
  {http://arxiv.org/abs/1311.5537} {arXiv:1311.5537 [hep-ph]} \BibitemShut
  {NoStop}%
\bibitem [{\citenamefont {Abel}\ \emph {et~al.}(2001)\citenamefont {Abel},
  \citenamefont {Khalil},\ and\ \citenamefont {Lebedev}}]{abel_edm_2001}%
  \BibitemOpen
  \bibfield  {author} {\bibinfo {author} {\bibfnamefont {S.}~\bibnamefont
  {Abel}}, \bibinfo {author} {\bibfnamefont {S.}~\bibnamefont {Khalil}}, \ and\
  \bibinfo {author} {\bibfnamefont {O.}~\bibnamefont {Lebedev}},\ }\href
  {\doibase 10.1016/S0550-3213(01)00233-4} {\bibfield  {journal} {\bibinfo
  {journal} {Nuclear Physics B}\ }\textbf {\bibinfo {volume} {606}},\ \bibinfo
  {pages} {151} (\bibinfo {year} {2001})},\ \bibinfo {note} {arXiv:
  hep-ph/0103320}\BibitemShut {NoStop}%
\bibitem [{\citenamefont {Ibrahim}\ and\ \citenamefont
  {Nath}(2008)}]{ibrahim_cp_2008}%
  \BibitemOpen
  \bibfield  {author} {\bibinfo {author} {\bibfnamefont {T.}~\bibnamefont
  {Ibrahim}}\ and\ \bibinfo {author} {\bibfnamefont {P.}~\bibnamefont {Nath}},\
  }\href {\doibase 10.1103//RevModPhys.80.577} {\bibfield  {journal} {\bibinfo
  {journal} {arXiv:0705.2008 [hep-ph]}\ } (\bibinfo {year} {2008}),\
  10.1103//RevModPhys.80.577},\ \bibinfo {note} {arXiv: 0705.2008}\BibitemShut
  {NoStop}%
\bibitem [{\citenamefont {Barr}\ and\ \citenamefont
  {Zee}(1990)}]{Barr_zee_edm}%
  \BibitemOpen
  \bibfield  {author} {\bibinfo {author} {\bibfnamefont {S.~M.}\ \bibnamefont
  {Barr}}\ and\ \bibinfo {author} {\bibfnamefont {A.}~\bibnamefont {Zee}},\
  }\href {\doibase 10.1103/PhysRevLett.65.21} {\bibfield  {journal} {\bibinfo
  {journal} {Phys. Rev. Lett.}\ }\textbf {\bibinfo {volume} {65}},\ \bibinfo
  {pages} {21} (\bibinfo {year} {1990})},\ \bibinfo {note} {[Erratum:
  Phys.Rev.Lett. 65, 2920 (1990)]}\BibitemShut {NoStop}%
\bibitem [{\citenamefont {Wu}\ \emph {et~al.}(2020)\citenamefont {Wu},
  \citenamefont {Han}, \citenamefont {Chow}, \citenamefont {Ang}, \citenamefont
  {Meisenhelder}, \citenamefont {Panda}, \citenamefont {West}, \citenamefont
  {Gabrielse}, \citenamefont {Doyle},\ and\ \citenamefont
  {DeMille}}]{ACME_next_gen}%
  \BibitemOpen
  \bibfield  {author} {\bibinfo {author} {\bibfnamefont {X.}~\bibnamefont
  {Wu}}, \bibinfo {author} {\bibfnamefont {Z.}~\bibnamefont {Han}}, \bibinfo
  {author} {\bibfnamefont {J.}~\bibnamefont {Chow}}, \bibinfo {author}
  {\bibfnamefont {D.~G.}\ \bibnamefont {Ang}}, \bibinfo {author} {\bibfnamefont
  {C.}~\bibnamefont {Meisenhelder}}, \bibinfo {author} {\bibfnamefont {C.~D.}\
  \bibnamefont {Panda}}, \bibinfo {author} {\bibfnamefont {E.~P.}\ \bibnamefont
  {West}}, \bibinfo {author} {\bibfnamefont {G.}~\bibnamefont {Gabrielse}},
  \bibinfo {author} {\bibfnamefont {J.~M.}\ \bibnamefont {Doyle}}, \ and\
  \bibinfo {author} {\bibfnamefont {D.}~\bibnamefont {DeMille}},\ }\href
  {\doibase 10.1088/1367-2630/ab6a3a} {\bibfield  {journal} {\bibinfo
  {journal} {New J. Phys.}\ }\textbf {\bibinfo {volume} {22}},\ \bibinfo
  {pages} {023013} (\bibinfo {year} {2020})},\ \Eprint
  {http://arxiv.org/abs/1911.03015} {arXiv:1911.03015 [physics.atom-ph]}
  \BibitemShut {NoStop}%
\bibitem [{\citenamefont {Li}\ \emph {et~al.}(2008)\citenamefont {Li},
  \citenamefont {Profumo},\ and\ \citenamefont {Ramsey-Musolf}}]{Li:2008kz}%
  \BibitemOpen
  \bibfield  {author} {\bibinfo {author} {\bibfnamefont {Y.}~\bibnamefont
  {Li}}, \bibinfo {author} {\bibfnamefont {S.}~\bibnamefont {Profumo}}, \ and\
  \bibinfo {author} {\bibfnamefont {M.}~\bibnamefont {Ramsey-Musolf}},\ }\href
  {\doibase 10.1103/PhysRevD.78.075009} {\bibfield  {journal} {\bibinfo
  {journal} {Phys. Rev. D}\ }\textbf {\bibinfo {volume} {78}},\ \bibinfo
  {pages} {075009} (\bibinfo {year} {2008})},\ \Eprint
  {http://arxiv.org/abs/0806.2693} {arXiv:0806.2693 [hep-ph]} \BibitemShut
  {NoStop}%
\bibitem [{\citenamefont {Hill}\ and\ \citenamefont
  {Solon}(2014)}]{hill_wimp-nucleon_2014}%
  \BibitemOpen
  \bibfield  {author} {\bibinfo {author} {\bibfnamefont {R.~J.}\ \bibnamefont
  {Hill}}\ and\ \bibinfo {author} {\bibfnamefont {M.~P.}\ \bibnamefont
  {Solon}},\ }\href {\doibase 10.1103/PhysRevLett.112.211602} {\bibfield
  {journal} {\bibinfo  {journal} {Physical Review Letters}\ }\textbf {\bibinfo
  {volume} {112}},\ \bibinfo {pages} {211602} (\bibinfo {year} {2014})},\
  \bibinfo {note} {arXiv: 1309.4092}\BibitemShut {NoStop}%
\end{thebibliography}%

\end{document}